\documentclass[aps,pra,floatfix,a12]{revtex4-2}
\usepackage{amsfonts}
\usepackage{amsmath}
\usepackage{amsthm}
\usepackage{amssymb}
\usepackage{graphicx}
\usepackage{appendix}
\usepackage{hyperref}
\usepackage{textcomp}
\usepackage{multirow}
\usepackage{color}
\usepackage{bm}
\usepackage{braket}
\usepackage{subfigure}

\begin{document}

\title{Multidimensional Soliton Systems}
\author{Boris A. Malomed}
\affiliation{Department of Physical Electronics, School of Electrical Engineering, Faculty of Engineering,
Tel Aviv University, Tel Aviv 69978, Israel\\
Instituto de Alta Investigaci\'{o}n, Universidad de Tarapac\'{a}, Casilla
7D, Arica, Chile\footnote}

\begin{abstract}
This concise review aims to provide a summary of the most relevant recent
experimental and theoretical results for solitons, i.e., self-trapped bound
states of nonlinear waves, in two- and three-dimensional (2D and 3D) media.
In comparison with commonly known one-dimensional solitons, which are,
normally, stable modes, a challenging problem is the propensity of 2D and 3D
solitons to instability, caused by the occurrence of the critical or
supercritical wave collapse (catastrophic self-compression) in the same
spatial dimension. A remarkable feature of multidimensional solitons is
their ability to carry vorticity; however, 2D vortex rings and 3D vortex
tori are subject to strong splitting instability. Therefore, it is natural
to categorize the basic results according to physically relevant settings
which make it possible to maintain stability of fundamental
(non-topological) and vortex solitons against the collapse and splitting,
respectively. The present review is focused on schemes that were recently
elaborated in terms of Bose-Einstein condensates and similar photonic
setups. These are two-component systems with \textit{spin-orbit coupling},
and ones stabilized by the beyond-mean-field \textit{Lee-Huang-Yang effect}.
The latter setting has been implemented experimentally, giving rise to
stable self-trapped quasi-2D and 3D \textit{quantum droplets}.
\end{abstract}

\maketitle
\tableofcontents


\textbf{Acronyms}

\noindent 1D -- one-dimensional

\noindent \noindent 2D -- two-dimensional

\noindent 3D -- three-dimensional

\noindent BEC -- Bose-Einstein condensate

\noindent CQ -- cubic-quintic (nonlinearity))

\noindent FT -- flat-top (profiles of solitons and quantum droplets)

\noindent GP -- Gross-Pitaevskii (equation)

\noindent GS -- ground state

\noindent GVD -- group-velocity dispersion

\noindent LHY -- Lee-Huang-Yang (correction to the MF theory)

\noindent MF -- mean-field (approximation)

\noindent NLS -- nonlinear Schr\"{o}dinger (equation)

\noindent QD -- quantum droplet

\noindent SOC -- spin-orbit coupling

\noindent STOV -- spatiotemporal vortex

\noindent TS -- Townes soliton

\noindent VA -- variational approximation

\noindent VK -- Vakhitov-Kolokolov (stability criterion)

\noindent ZS -- Zeeman splitting

\newpage

\section*{Introduction: multidimensional solitons and their proneness to
instabilities}

The absolute majority of theoretical and experimental work which has been
performed for solitons, i.e., self-trapped solitary waves in nonlinear
systems, deal with one-dimensional (1D) settings. Extension of the soliton
concepts to multidimensional spaces is a very promising, but also really
challenging, direction. A fascinating possibility is to create new species
of solitary states in the two- and three-dimensional (2D and 3D) geometries
with intrinsic topological structures. An obvious option is to create 2D and
3D \textit{vortex solitons} carrying an intrinsic angular momentum, which
may be considered as a classical counterpart of the spin of quantum
particles. Multi-component solitons can be used to build more sophisticated
topological structures, such as famous skyrmions, hopfions (alias twisted
vortex tori in the 3D space), monopoles, knots, and others \cite{Makhankov
book,Manton-Satcliffe-nucl-phys,Manton-book,Radu Volkov,Yasha}.

However, the work with 2D and 3D solitons encounters fundamental
difficulties. First, unlike the fundamental 1D models that give rise to
solitons, such as the Korteweg - de Vries, since-Gordon, and nonlinear Schr%
\"{o}dinger (NLS) equations, multidimensional equations are not integrable
(there are some integrable 2D models -- most notably, the
Kadomtsev-Petviashvili equations \cite{Biondini Pelinovsky}). The lack of
the integrability of basic 2D and 3D soliton equations may be considered as
a technical difficulty, because relevant solutions, once they are not
available in an exact form, can be constructed by means of approximate
analytical methods, such as the variational approximation (VA), and the
availability of powerful computers makes it possible to produce numerical
solutions in all relevant cases. However, in 2D and 3D settings,
self-attractive nonlinearities, which are necessary for the formation of
solitons give rise to a fundamental problem, \textit{viz}., instability of
the localized states, due to the fact that the same settings give rise to
the \textit{wave collapse} (alias blowup, i.e., catastrophic
self-compression of the wave field leading to the formation of a true
singularity after a finite evolution time), which is illustrated by Fig. \ref%
{fig1}(a) \cite{Berge',Sulem,ZakhKuz,Fibich}.
\begin{figure}[tbp]
\begin{center}
\subfigure[]{\includegraphics[width=0.45\textwidth]{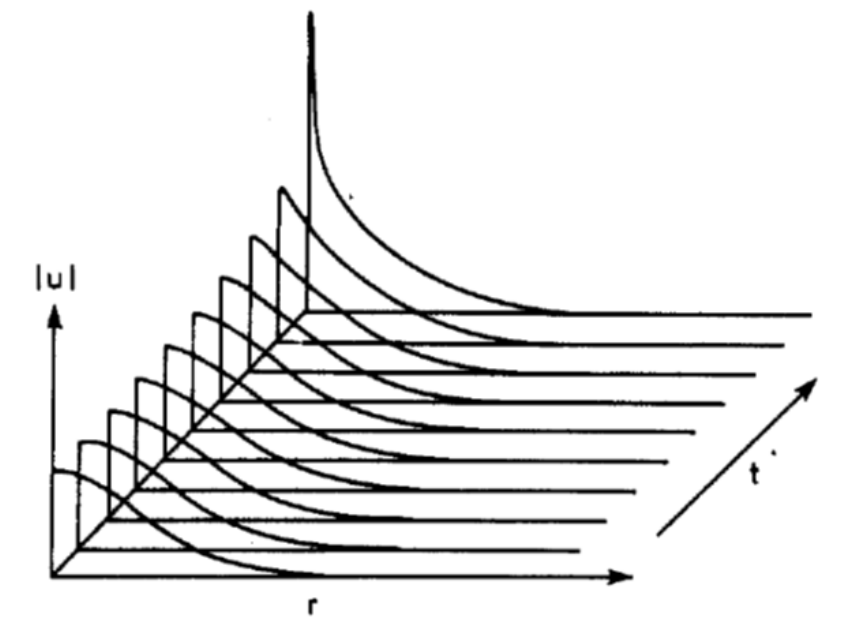}} %
\subfigure[]{\includegraphics[width=0.70\textwidth]{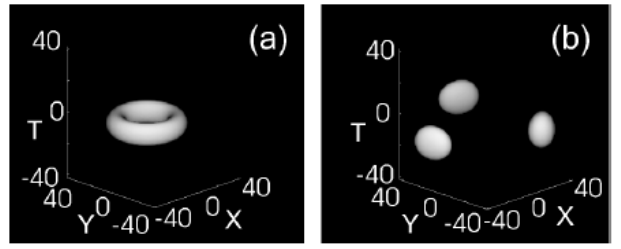}}
\end{center}
\caption{(a) Onset of the critical collapse featured by the unstable
evolution of a TS (Townes soliton), as produced by simulations of the 2D
version of Eq. (\protect\ref{NLS}) with $\protect\sigma =3$, shown in the
radial cross-section. (b) An example of the splitting instability of a 3D
soliton with embedded vorticity $S=2$ (a vortex torus). The left and right
panels display surfaces $\left\vert u\left( x,y,\protect\tau ,z\right)
\right\vert =\mathrm{const}$, which are produced, respectively, as a
stationary solution of the 3D counterpart of Eq. (\protect\ref{2D NLS}) (the
input for simulations of the full equation), and as a result of the simuted
evolution. The figures are borrowed from book \protect\cite{book}.}
\label{fig1}
\end{figure}

The instability problem is exhibited by the general model based on the NLS
equation in space of dimension $D$ with the self-attractive nonlinear term
of power $\sigma $ [the Kerr nonlinearity in optics \cite{KiAgr} and
collisional nonlinearity in Bose-Einstein condensates (BECs) \cite{Pit-Str}
correspond to $\sigma =3$]:%
\begin{equation}
i\psi _{t}+\frac{1}{2}\nabla ^{2}\psi +|\psi |^{\sigma -1}\psi =0.
\label{NLS}
\end{equation}%
Complex field $\psi $ in Eq. (\ref{NLS}) represents the local amplitude of
electromagnetic waves in optics, or the wave function of \textit{matter waves%
} in BEC, considered in the mean-field (MF) approximation. Solutions of Eq. (%
\ref{NLS}) are characterized by the dynamical invariants, \textit{viz}., the
norm%
\begin{equation}
N=\int \left\vert \psi \left( x,y,z\right) \right\vert ^{2}d^{D}x,  \label{N}
\end{equation}%
and Hamiltonian,%
\begin{equation}
H=\int \left( \frac{1}{2}|\nabla \psi |^{2}-\frac{2}{\sigma +1}|\psi
|^{\sigma +1}\right) d^{D}x\equiv H_{\mathrm{grad}}+H_{\mathrm{self-focusing}%
},  \label{HHH}
\end{equation}%
where the integration is performed over the $D$-dimensional space with
coordinates $\{x_{1},~...~,x_{D}\}$.

In optics, the spatiotemporal propagation of light in media with \emph{%
anomalous} group-velocity dispersion (GVD) is governed by the NLS equation (%
\ref{NLS}), in which time $t$ is replaced by the propagation distance, $z$,
while the temporal variable,
\begin{equation}
\tau \equiv t-z/V_{\mathrm{gr}},  \label{tau}
\end{equation}%
where $V_{\mathrm{gr}}$ is the group velocity of the carrier wave \cite%
{KiAgr}, plays the role of an additional coordinate. Spatiotemporal solitons
predicted by such a model are often called \textit{light bullets} (LBs) \cite%
{Silberberg}.

Stationary solutions to Eq. (\ref{NLS}) are looked for as%
\begin{equation}
\psi (x_{1},~...~,x_{D};t)=e^{-i\mu t}\phi (x_{1},~...~,x_{D}),  \label{mu}
\end{equation}%
where real constant $\mu $ is called the chemical potential in the context
of matter waves in BEC \cite{Pit-Str}, and stationary wave function $\phi $
is determined by equation%
\begin{equation}
\mu \phi +\frac{1}{2}\nabla ^{2}\phi +|\phi |^{\sigma -1}\phi =0  \label{phi}
\end{equation}%
(localized solutions of Eq. (\ref{phi}) may only corresponds to $\mu <0$).

To explain the onset of the collapse in the framework of Eq. (\ref{NLS}),
one can consider, following Ref. \cite{ZakhKuz}, a localized isotropic
configuration of field $\psi $, with amplitude $A$ and size (radius) $R$. An
obvious estimate for the norm is%
\begin{equation}
N\sim A^{2}R^{D}.  \label{N est}
\end{equation}%
Similarly, the gradient and self-focusing terms in Hamiltonian (\ref{HHH})
are estimated as follows, eliminating $A^{2}$ in favor of $N$ by means of
Eq. (\ref{N est}):%
\begin{equation}
H_{\mathrm{grad}}\sim NR^{-2},H_{\mathrm{self-focusing}}\sim -N^{(\sigma
+1)/2}R^{-(\sigma -1)D/2}.~  \label{HH}
\end{equation}%
The collapse, i.e., catastrophic shrinkage of the state towards $%
R\rightarrow 0$, takes place if the consideration of $R\rightarrow 0$ for
fixed $N$ reveals that $H(R\rightarrow 0)\rightarrow -\infty $ (in other
words, the system's ground state (GS) formally corresponds to $R\rightarrow 0
$). The comparison of the two terms in Eq. (\ref{HH}) demonstrates that the
unconditional (alias \textit{supercritical}) collapse, which occurs if $%
\left\vert H_{\mathrm{self-focusing}}\right\vert $ diverges at $R\rightarrow
0$ faster than $H_{\mathrm{grad}}$, takes place under the condition of%
\begin{equation}
\left( \sigma -1\right) D>4.  \label{>4}
\end{equation}%
In this case, the collapse is initiated by the input with an arbitrarily
small norm $N$. In the \textit{critical case}, which corresponds to
\begin{equation}
\left( \sigma -1\right) D=4  \label{=4}
\end{equation}%
(in particular, Eq. (\ref{=4}) holds for the 2D cubic ($\sigma =3$) NLS
equation, as well as for the 1D equation with the quintic nonlinearity, $%
\sigma =5$ \cite{1D Townes}), both terms in Eq. (\ref{HH}) feature the same
scaling at $R\rightarrow 0$, the \textit{critical collapse} taking place if $%
N$ exceeds a certain critical (threshold) value, $N_{\mathrm{cr}}$. In the
case of Eq. (\ref{NLS}) with $D=2$ and $\sigma =3$,
the numerically found critical norm is%
\begin{equation}
N_{\mathrm{cr}}\approx 5.85.  \label{Pcr}
\end{equation}%
In an analytical form, the critical norm was predicted by the VA, based on
the Gaussian ansatz, with a relative error $\simeq 7\%$ \cite{Desaix}:%
\begin{equation}
\left( N_{\mathrm{cr}}\right) _{\mathrm{VA}}=2\pi .  \label{PcrVA}
\end{equation}%
The collapse does not occur, hence the solitons produced by Eq. (\ref{NLS})
may be stable, in the case of
\begin{equation}
\left( \sigma -1\right) D<4  \label{<4}
\end{equation}%
(for instance, the classical 1D solitons with the cubic nonlinearity \cite%
{Zakharov-Shabat}).

Condition (\ref{<4}) also holds in all physically relevant dimensions, $%
D\leq 3$, for the quadratic nonlinearity ($\sigma =2$), even if the
quadratic nonlinearity is realized not as a single NLS equation, but rather
in the form of two coupled equations for the fundamental and second
harmonics \cite{Kanashov,Hayata Koshiba,Turitsyn,Malomed 1997}. Accordingly,
stable 2D spatial \cite{Torruellas} and spatiotemporal \cite{Wise} solitons
were experimentally created in optical media with the
second-harmonic-generating nonlinearity.

For the same case of Eq. (\ref{NLS}) with $D=2$ and $\sigma =3$, the
occurrence of the critical collapse is a consequence of the exact \textit{%
virial theorem} \cite{Vlasov}, which is a corollary of Eq. (\ref{NLS}):%
\begin{equation}
\frac{d^{2}}{dt^{2}}\left( N\left\langle r^{2}\right\rangle \right) \equiv
\frac{d^{2}}{dt^{2}}\int \int r^{2}|\psi |^{2}dxdy=2H,  \label{virial}
\end{equation}%
where $H$ is Hamiltonian (\ref{H}) for $D=2$ and $\sigma =3$, $N$ is norm (%
\ref{N}), and $\left\langle r^{2}\right\rangle $ is the mean value of the
squared radius of the localized configuration of the wave function. Because $%
N$ and $H$ are dynamical invariants (constants), a solution of Eq. (\ref%
{virial}) gives%
\begin{equation}
\left\langle r^{2}\right\rangle (t)=\left\langle r^{2}\right\rangle
(t=0)+Ct+\left( H/N\right) t^{2},  \label{virial2}
\end{equation}%
with a constant $C$. Thus, $H>0$ implies $\left\langle r^{2}\right\rangle
\rightarrow \infty $ at $t\rightarrow \infty $, i.e., decay of the wave field%
$.$ On the other hand, in the case of $H<0$ Eq. (\ref{virial2}) shows that
the mean radius vanishes as
\begin{equation}
\left\langle r^{2}\right\rangle \sim \left( t_{\mathrm{cr}}-t\right)
\label{tcr}
\end{equation}%
at some critical time $t_{\mathrm{cr}}$. Then, the conservation of the norm
suggests that the amplitude of the field diverges as
\begin{equation}
\left\vert \psi \right\vert _{\max }\sim \left( t_{\mathrm{cr}}-t\right)
^{-1/2}  \label{psimax}
\end{equation}%
at $t\rightarrow t_{\mathrm{cr}}$, signaling the emergence of the
singularity.

The asymptotic stage of the supercritical collapse, governed by the 3D
version of Eq. (\ref{NLS}) with $\sigma =3$, is different, featuring $%
\left\langle r^{2}\right\rangle \sim \left( t_{\mathrm{cr}}-t\right) ^{4/5}$
and $\left\vert \psi \right\vert _{\max }\sim \left( t_{\mathrm{cr}%
}-t\right) ^{-3/5}$, instead of Eqs. (\ref{tcr}) and (\ref{psimax}),
respectively. These conclusions can be obtained in an analytical form by
means of the VA \cite{ZakhKuz}.

Thus, in the case when condition (\ref{>4}) holds, small perturbations added
to any soliton of the NLS equation trigger its blowup, while in the critical
case, defined by Eq. (\ref{=4}), small perturbations initiate the
instability in the form of either the blowup or decay (i.e., the respective
solitons represent a \textit{separatrix} between collapsing and decaying
solutions of the 2D NLS equation, with $N<N_{\mathrm{cr}}$ and $N>N_{\mathrm{%
cr}}$, respectively). In this connection, it is relevant \ to note that the
first species\emph{\ }of solitons which was\emph{\ }ever considered in
optics is the family of \textit{Townes solitons} (TSs) \cite{Townes}. These
are stationary self-trapped solutions of the 2D version of Eq. (\ref{NLS}%
) with the cubic nonlinearity ($\sigma =3$), which form a \textit{degenerate
family}, as norm (\ref{N}) takes the single value, $N=N_{\mathrm{cr}}$ (the
same one as given by Eq. (\ref{Pcr})), for the entire family. The norm
degeneracy of the TS family is explained by the fact Eq. (\ref{phi}) is
invariant with respect to rescaling,%
\begin{equation}
\mu \rightarrow l^{2}\mu ,~\left( x_{1},...x_{D}\right) \rightarrow
l^{-1}\left( x_{1},...x_{D}\right) ,~u\rightarrow l^{2/(\sigma -1)}u,~
\label{scaling}
\end{equation}%
with arbitrary factor $l$. The invariance makes it possible to generate the
entire family from a single soliton, the respective rescaling of the norm
being%
\begin{equation}
N\rightarrow l^{-D+4/(\sigma -1)}N,  \label{Pscaling}
\end{equation}%
where $D$ is the dimension. The conclusion is that all the solitons indeed
have the same norm, according to Eq. (\ref{Pscaling}), exactly in the
critical case defined by Eq. (\ref{=4}), including the TSs corresponding to $%
D=2$, $\sigma =3$.

Note that the same analysis produces a scaling relation between $\mu $ and $N
$ for soliton families generated by Eq. (\ref{NLS}):%
\begin{equation}
N\sim \left( -\mu \right) ^{-D/2+2/(\sigma -1)}  \label{N-mu}
\end{equation}%
($-\mu $ is written here as only $\mu <0$ may correspond to solitons, as
mentioned above). Relation (\ref{N-mu}) makes it possible to apply the
celebrated Vakhitov-Kolokolov (VK) criterion,
\begin{equation}
dN/d\mu <0,  \label{dN/dmu}
\end{equation}%
which is a necessary stability condition for any soliton family maintained
by the self-attraction, irrespective of the spatial dimension \cite%
{Vakh-Kol,Berge',ZakhKuz}. Indeed, the application of the VK criterion to
relation (\ref{N-mu}) demonstrates that the solitons may only be stable in
exactly the same case (\ref{<4}) when the possibility of the collapse is
eliminated

TSs have never been observed experimentally in optics, because, as said
above, they are destabilized by the critical collapse. Nevertheless,
observations of carefully engineered matter-wave TSs in 2D binary BECs of
ultracold atomic gases, for which the respective Gross-Pitaevskii (GP)
equation is essentially the same as the 2D equation (\ref{NLS}) with $\sigma
=3$, were reported recently \cite{Chen Hung,Townes sol 2-comp Rb-87,Chen
Hung2}. These observations are possible because the initial stage of the
growth of the instability driven by the critical collapse is \emph{%
subexponential} (i.e., very slow).

3D and 2D solitons with embedded vorticity (alias \textit{vortex tori} and
\textit{vortex rings}, respectively) are looked for as solutions of Eq. (\ref%
{NLS}) written in terms of cylindrical coordinates $\left( \rho ,\theta
,z\right) $ (in the 2D case, $z$ is dropped),%
\begin{equation}
\psi =e^{-i\mu t+iS\theta }\phi \left( \rho ,z\right) ,  \label{rho}
\end{equation}%
cf. Eq. (\ref{tau}), where the integer \textit{winding number} (alias
topological charge) $S$ represents the vorticity, and real function $\phi $
satisfies the equation%
\begin{equation}
\mu \phi +\frac{1}{2}\left( \frac{\partial ^{2}}{\partial \rho ^{2}}+\frac{1%
}{\rho }\frac{\partial }{\partial \rho }-\frac{S^{2}}{\rho }+\frac{\partial
^{2}}{\partial z^{2}}\right) \phi +\phi ^{\sigma }=0.  \label{rho-real}
\end{equation}%
The vortex states are naturally related to the $z$-component of the angular
momentum,%
\begin{equation}
M=i\int \left( y\frac{\partial \psi }{\partial x}-x\frac{\partial \psi }{%
\partial y}\right) d^{D}x,  \label{M}
\end{equation}%
which is a dynamical invariant of Eq. (\ref{NLS}). For the stationary
vortex-soliton states, the angular momentum is simply related to norm
(\ref{N}), \textit{viz}., $M=SN$.

It is relevant to mention that gradually expanding spatiotemporal pulses
with toroidal shapes were experimentally created in linear optics \cite%
{Ellenbogen,Zayats}, although these are not self-trapped modes.

In terms of the optical spatiotemporal propagation, with $t$ replaced by $z$
and the additional coordinate being the temporal one (\ref{tau}), the
orientation of the corresponding vector $\mathbf{M}$ of the conserved
angular momentum may be arbitrary in the 3D space $\left( x,y,\tau \right) $%
. This option implies the possibility of the creation of \textit{%
spatiotemporal optical vortices} (STOVs), if vector $\mathbf{M}$ in the 3D
space is not aligned with axis $\tau $. This possibility was first explored
in Ref. \cite{Dror spatiotemp vort} in a 2D form, considering the
spatiotemporal light propagation in a planar waveguide with transverse
coordinate $x$ and the intrinsic cubic-quintic (CQ) nonlinearity. The
respective 2D NLS equation for the slowly varying amplitude $u\left( x,\tau
,z\right) $ of the electromagnetic wave, written in the scaled form, is%
\begin{equation}
iu_{z}+\frac{1}{2}\left( u_{xx}+u_{\tau \tau }\right) +|u|^{2}u-|u|^{4}u=0.
\label{2D NLS}
\end{equation}%
The STOV solutions to Eq. (\ref{2D NLS}), with real propagation constant $k>0
$ and integer vorticity $S$, can be looked for as%
\begin{equation}
u\left( x,\tau ,z\right) =e^{ikz+iS\theta }U(r),  \label{uU}
\end{equation}%
where $r\equiv \sqrt{x^{2}+\tau ^{2}}$ and $\theta \equiv \arctan \left(
\tau /x\right) $ are the \textit{spatiotemporal} radial and angular
coordinates, and real amplitude function $U$ satisfies the equation%
\begin{equation}
kU=\frac{1}{2}\left( \frac{d^{2}}{dr^{2}}+\frac{1}{r}\frac{d}{dr}-\frac{S^{2}%
}{r^{2}}\right) U+U^{3}-U^{5}.  \label{UCQ}
\end{equation}%
As proposed in Ref. \cite{Dror spatiotemp vort}, STOVs in the planar
waveguide can be excited by an oblique external vortex beam. Later, the
concept of STOV was experimentally realized \cite{Milchberg,Chong} and
theoretically developed \cite{Bliokh STOV,Porras STOV} in the 3D form.

Vortex solitons may be subject to the strong instability which develops
faster than the collapse, \textit{viz}., spontaneous splitting of the\
vortex torus or ring in two or several fragments, which are close to the
corresponding fundamental (zero-vorticity) solitons, see an illustration in
Fig. \ref{fig1}(b) \cite{PhysD-vortices}. At a later stage of the evolution,
the secondary solitons may be destroyed by the collapse. In particular, in
the case of the 2D version of Eq. (\ref{NLS}) with $\sigma =3$, one can
construct families of TSs with embedded vorticity $S$ \cite{TS with S > 0}.
For each integer $S$, the family is degenerate, similar to the one with $S=0$%
, in the sense that all the vortex TSs have the single value, $N_{\mathrm{cr}%
}^{(S)}$, of their norm, cf. Eq. (\ref{Pcr}). For $S\geq 1$, the dependence
of the critical value on $S$ is predicted, in a rather accurate form, by an
analytical approximation derived in Ref. \cite{Dong giant vort}: $N_{\mathrm{%
cr}}^{(S)}\approx 4\sqrt{3}\pi S$.

Even if the collapse is eliminated, e.g., in the case of the saturable
self-focusing nonlinearity, vortex solitons may remain unstable against
spontaneous splitting \cite{Skryabin}.

2D and 3D versions of Eq. (\ref{NLS}), especially the ones with the cubic
nonlinearity, $\sigma =3$, are basic models for many physical realizations
in optics, BECs (matter waves), plasmas (Langmuir waves), etc., but the
collapse-driven instability implies that these physical settings cannot be
used straightforwardly for the creation of multidimensional solitons.
Therefore, a\ cardinal problem is search for physically relevant
multidimensional systems which include additional ingredients that make it
possible to suppress the collapse and produce stable 2D and 3D solitons,
both fundamental and vortical ones. This is indeed possible in various
physical setups. In particular, stable 2D and 3D optical solitons can be
predicted if the optical medium features, in addition to the cubic
self-focusing, quintic \emph{self-defocusing} (see Eq. (\ref{2D NLS}))\ or,
more generally, \emph{saturation}, that arrests the blowup and indeed
provides the full stabilization of 2D and 3D fundamental solitons and
partial stabilization and vortical 2D and 3D ones (2D fundamental solitons
stabilized by the quintic self-defocusing have been reported experimentally
\cite{Edilson}, and partial stabilization of 2D solitary vortices in a
saturable medium was reported too \cite{Cid-vortex}). A recently discovered
option is to consider a binary BEC with the attractive cubic interaction
between its two components, which is slightly stronger that the intrinsic
self-repulsion in each component. In this system, the collapse is arrested
by a higher-order \emph{quartic} self-repulsive term, which is induced by
the \textit{Lee-Huang-Yang} (LHY) effect, i.e., a correction to the usual
cubic MF interaction induced by quantum fluctuations around the MF state
\cite{LHY}. As a result, the binary BEC creates completely stable 3D and
quasi-2D self-trapped \textquotedblleft quantum droplets" (QDs), which seem
as multidimensional solitons (even if\ they are not usually called
\textquotedblleft solitons", as the name of QDs is commonly used). The
prediction of QDs \cite{Petrov,Petrov Astrakharchik} was quickly realized
experimentally \cite{Leticia1,Inguscio,hetero QD}. It was also predicted,
although not yet demonstrated experimentally, that 2D \cite{2D vortex QD}
and 3D \cite{swirling} QDs with embedded vorticity may be stable too.

Various aspects of theoretical and experimental studies of multidimensional
solitons were subjects of several review articles \cite{Malomed 2005,nonlin
latt RMP,Malomed 2016,Special Topics,PhysD-vortices,Nature Phys Rev,Sandy
review,dissipa multidim}. Recently, the results have been summarized in a
comprehensive book \cite{book}. Physical realizations in which stable
multidimensional solitons can be created chiefly belong to two broad areas:
first, optics and photonics, and, second, various settings based on BECs.
This review article is focused on the newest and most perspective
theoretical predictions and experimental findings, \textit{viz}., (i) the
use of spin-orbit coupling (SOC) in binary BECs and its counterpart in
optics (Sections 2 and 3), and (ii) the prediction and experimental
realization of stable quasi-2D and 3D localized states in the form of QDs
(Section 4).

\section{Stabilization of 2D and 3D matter-wave solitons by the spin-orbit
coupling (SOC)}

\subsection{Introduction to the topic}

The use of matter waves in BEC for emulation of fundamental effects from
condensed-matter physics is a well-known subject \cite{emulator}. While in
original settings those effect usually appear in a complex form, the
matter-wave emulation may offer a much simpler possibility to reproduce
their basic features. In particular,\ much interest has been drawn to the
emulation of SOC in a binary (two-component) BEC. Although the true spin of
bosonic atoms, such as $^{87}\mathrm{Rb}$, used for the creation of BEC in
ultracold gases, is zero, the wave function of a binary condensate, which is
a mixture of two different hyperfine atomic states, may be considered as a
two-component \textit{pseudo-spinor}, which corresponds to \textit{pseudospin%
} $1/2$. In its original form, SOC originates in physics of semiconductors,
as the weakly-relativistic interaction between the electron's magnetic
moment and its motion through the electrostatic field of the ionic lattice.
Mapping the electron's spinor wave function into the pseudo-spinor bosonic
wave function of the binary condensate, SOC is emulated by the linear
interaction between the pseudospin and motion of the bosonic atoms \cite%
{Spielman-Nature,SOC in BEC}. The Zeeman splitting (ZS) between the two
components of the binary BEC may also play an important role in this setting
\cite{Zeeman}.

Two fundamental types of the SOC, which are known in physics of
semiconductors, are represented by the Dresselhaus \cite{Dresselhaus} and
Rashba \cite{Rashba} Hamiltonians. Both these types can be emulated in
atomic BECs. A majority of experimental works were dealing with effectively
one-dimensional SOC settings. Nevertheless, the realization of SOC in the 2D
BEC was reported too \cite{2D SOC}, which makes it relevant to consider 2D
and, eventually, 3D SOC\ states, including multidimensional solitons.

While the SOC in BEC is a linear effects, its interplay with the intrinsic
self-repulsive BEC nonlinearity was predicted to produce a variety of
dynamical states, such as vortices \cite{SOC-vort1,SOC-vort Pu,SOC-vort2},
monopoles \cite{SOC-monopoles}, skyrmions \cite{SOC-skyrme1,SOC-skyrme2},{\
etc.} Further, solitons in binary condensates with the cubic \emph{%
self-attractive} intrinsic nonlinearity may be essentially affected by SOC.
A surprising result is that two different families of 2D solitons, namely
\textit{semi-vortices} (SVs, with winding numbers (topological charges) $m=0$
and $\pm 1$ separately carried by the two components), and \textit{mixed
modes} (MMs, which combine $m=\left( 0,-1\right) $ and $m=(0,+1)$ in the
different components) become \emph{stable} in the binary system with the
linear SOC of the Rashba type \cite{Ben Li,Luca-Wesley-SOC,Sak-Mal}, as well
as under the action of the combined Rashba-Dresselhaus SOC \cite%
{Sherman,Romanian-Sherman}. Furthermore, the SV (MM) solitons realize the GS
of the 2D binary system when the self-attraction in two components of the
wave function is stronger (weaker) than the cross-attraction between them.
Prior to reporting these findings, it was commonly believed that 2D systems
of the NLS/GP type with cubic self-attraction could never give rise to
stable solitons, and did not have a GS, because of the occurrence of the
critical collapse in the same system (peculiarities of the onset of the
critical collapse in the 2D SOC system were studied in Ref. \cite{Mardonov}).

The stabilizing effect of the SOC for the 2D solitons in free space is
underlain by the fact that SOC sets up its length scale, which is inversely
proportional to the SOC strength. The fixed scale breaks the above-mentioned
scale invariance of the 2D NLS equation (see Eqs. (\ref{scaling}) and (\ref%
{Pscaling})), thus lifting the norm degeneracy of the solitons, and pushing
their norm \emph{below} the threshold necessary for the onset of the
critical collapse. Thus, getting protected against the critical collapse,
the solitons retrieve their stability and realize the GS \cite{Ben Li}.

In the 3D system, characterized by the \emph{supercritical collapse} (see
above), the same protection mechanism does not work. Nevertheless, the
interplay of the linear SOC and cubic self-attraction produces 3D \textit{%
metastable solitons} of the same two types, SV and MM, which are stable
against small perturbations but can develop the collapse as a result of a
suddenly applied strong compression \cite{Han Pu 3D}.

\subsection{The 2D SOC system: semi-vortices (SVs) and mixed modes (MMs)}

\subsubsection{The general model and its bandgap spectra}

In the MF approximation, the 2D binary condensate is described by a
two-component wave function, $\left( \phi _{+},\phi _{-}\right) $, with
total norm%
\begin{equation}
N=\iint (|\phi _{+}|^{2}+|\phi _{-}|^{2})dxdy\equiv N_{+}+N_{-},
\label{Nsoc}
\end{equation}%
which is proportional to the total number of atoms in the condensate. The
evolution of the wave function is governed by the system of coupled GP
equations, written here in the scaled form (Dalibard \textit{et al}. 2011):%
\begin{equation}
\hspace{-11mm}i\frac{\partial \phi _{+}}{\partial t}=-\frac{1}{2}\nabla
^{2}\phi _{+}-(|\phi _{+}|^{2}+\gamma |\phi _{-}|^{2})\phi _{+}+\left(
\lambda \widehat{D}^{[-]}\phi _{-}-i\lambda _{D}\widehat{D}^{[+]}\phi
_{-}\right) -\Omega \phi _{+},  \label{SOCphi-}
\end{equation}%
\begin{equation}
\hspace{-11mm}i\frac{\partial \phi _{-}}{\partial t}=-\frac{1}{2}\nabla
^{2}\phi _{-}-(|\phi _{-}|^{2}+\gamma |\phi _{+}|^{2})\phi _{-}-\left(
\lambda \widehat{D}^{[+]}\phi _{+}+i\lambda _{D}\widehat{D}^{[-]}\phi
_{+}\right) +\Omega \phi _{-},  \label{SOCphi+}
\end{equation}%
where the nonlinear interactions are attractive, $\gamma $ is the relative
strength of the cross-attraction between the components, while the
self-attraction strength is scaled to be $1$. Further, $\lambda $ and $%
\lambda _{D}$ are constants of the Rashba and Dresselhaus SOC, respectively,
and $\widehat{D}^{[\pm ]}\equiv \partial /\partial x\pm i\partial /\partial y
$. The ZS terms in Eqs. (\ref{SOCphi-}) and (\ref{SOCphi+}) with strength $%
\Omega >0$, which lift the symmetry between the components $\phi _{\pm }$ of
the wave function, may be induced by a synthetic magnetic field directed
along the $z$ axis \cite{Zeeman}.

The spectrum of plane waves generated by the linearized version of Eqs. (\ref%
{SOCphi-}) and (\ref{SOCphi+}), $\phi _{\pm }\sim \exp \left( i\mathbf{k}%
\cdot \mathbf{r}-i\mu _{\pm }t\right) $, where $\mathbf{k}=\left(
k_{x},k_{y}\right) $ is the wave vector, contains two branches:
\begin{equation}
\mu _{\pm }=\frac{k^{2}}{2}\pm \sqrt{(\lambda ^{2}+\lambda
_{D}^{2})k^{2}+4\lambda \lambda _{D}k_{x}k_{y}+\Omega ^{2}}.
\label{SOC spectrum}
\end{equation}%
Solitons, in the form of $\phi _{1,2}\left( x,y,t\right) =e^{-i\mu t}\Phi
_{1,2}\left( x,y\right) $, may exist with values of the chemical potential $%
\mu $ belonging to the \textit{semi-infinite bandgap}, which is not covered
by expressions (\ref{SOC spectrum}) with $-\infty <k_{x},k_{y}<+\infty $,
i.e., at%
\begin{equation}
\mu <\mu _{\min }=%
\begin{array}{c}
-\frac{1}{2}\left[ \left( \lambda +\lambda _{D}\right) ^{2}+\Omega
^{2}\left( \lambda +\lambda _{D}\right) ^{-2}\right] ,~\text{for~}\left(
\lambda +\lambda _{D}\right) ^{2}>\Omega , \\
-\Omega ,~\text{for }\left( \lambda +\lambda _{D}\right) ^{2}<\Omega ~.%
\end{array}
\label{semi-inf}
\end{equation}

In the limit case of strong SOC, one may neglect the kinetic-energy terms ($%
\sim \nabla ^{2}$) in Eqs. (\ref{SOCphi-}) and (\ref{SOCphi+}) in comparison
to the SOC ones. This approximation produces a completely different spectrum
of plane waves,%
\begin{equation}
\mu =\pm \sqrt{\Omega ^{2}+\left( \lambda k_{x}+\lambda _{D}k_{y}\right)
^{2}+\left( \lambda _{D}k_{x}+\lambda k_{y}\right) ^{2}}.  \label{gap}
\end{equation}%
This spectrum includes a \textit{finite bandgap} (provided that the system
includes the ZS),
\begin{equation}
-\Omega <\mu <+\Omega ,  \label{finite gap}
\end{equation}
which may be populated by \textit{gap solitons} \cite{SOC 2D gap sol
Hidetsugu}. A part of solutions for the gap solitons are stable for $\gamma
<\gamma _{\max }\approx 0.77$ ($\gamma $ is the same cross-attraction
coefficient as in Eqs. (\ref{SOCphi-}) and (\ref{SOCphi+})) \cite{SOC 2D gap
sol Hidetsugu}.

Strictly speaking, the finite bandgap (\ref{finite gap}) does not exist, in
terms of the full system of Eqs. (\ref{SOCphi-}) and (\ref{SOCphi+}), as it
is covered by the plane-wave spectrum (\ref{SOC spectrum}) at very large
values of the wavenumbers, $\left\vert k_{x.y}\right\vert \gtrsim \lambda
,\lambda _{D}.$ Accordingly, the gap solitons, as solutions of the full
system, will suffer exponentially weak decay into small-amplitude plane
waves (\textit{radiation}).

The total energy of the system of Eqs. (\ref{SOCphi-}) and (\ref{SOCphi+})
is the sum of the kinetic ($E_{\mathrm{k}}$), self-interaction ($E_{\mathrm{%
int}}$), SOC ($E_{\mathrm{soc}}$), and ZS ($E_{\mathrm{ZS}}$) terms:
\begin{equation}
E=E_{\mathrm{k}}+E_{\mathrm{int}}+E_{\mathrm{soc}}+E_{\mathrm{ZS}},
\label{EEEEE}
\end{equation}%
where
\begin{equation}
\hspace{-11mm}E_{\mathrm{k}}=\frac{1}{2}\iint \left( |\nabla \phi
_{+}|^{2}+|\nabla \phi _{-}|^{2}\right) dxdy,  \label{Ek}
\end{equation}%
\begin{equation}
\hspace{-11mm}E_{\mathrm{int}}=-\frac{1}{2}\iint \left[ \left( |\phi
_{+}|^{4}+|\phi _{-}|^{4}\right) +2\gamma |\phi _{+}|^{2}|\phi _{-}|^{2}%
\right] dxdy,  \label{Eint}
\end{equation}%
\begin{equation}
\hspace{-11mm}E_{\mathrm{soc}}=\iint \left[ \phi _{+}^{\ast }\left( \lambda
\widehat{D}^{[-]}-i\lambda _{D}\widehat{D}^{[+]}\right) \phi _{-}-\phi
_{-}^{\ast }\left( \lambda \widehat{D}^{[+]}+i\lambda _{D}\widehat{D}%
^{[-]}\right) \phi _{+}\right] dxdy,  \label{ESOC}
\end{equation}%
\begin{equation}
\hspace{-11mm}E_{\mathrm{ZS}}=-\Omega \iint \left( |\phi _{+}|^{2}-|\phi
_{-}|^{2}\right) dxdy.  \label{EZS}
\end{equation}%
Below, following Ref. \cite{Ben Li}, the most essential results are
presented for 2D solitons maintained by SOC of the Rashba type, setting $%
\lambda _{D}=0$ and $\Omega =0$ (neglecting ZS) in Eqs. (\ref{SOCphi-}) and (%
\ref{SOCphi+}).

\subsubsection{SV solitons}

First, the system admits the substitution of the ansatz in the form of the
zero-vorticity (fundamental) component in $\phi _{+}$ and vortex one in $%
\phi _{-}$, written in terms of the polar coordinates $\left( r,\theta
\right) $ instead of $\left( x,y\right) $:

\begin{equation}
\phi _{+}\left( x,y,t\right) =e^{-i\mu t}f_{1}(r^{2}),~\phi _{-}\left(
x,y,t\right) =e^{-i\mu t+i\theta }rf_{2}(r^{2}),  \label{ansatz}
\end{equation}%
where real functions $f_{1,2}(r^{2})$ take final values at $r=0$,
exponentially decaying at $r\rightarrow \infty $. Composite modes of this
type, with winding numbers $0$ and $1$ in components $\phi _{+}$ and $\phi
_{-}$, are called SVs. The invariance of Eqs. (\ref{SOCphi-}) and (\ref%
{SOCphi+}), with respect to the transformation
\begin{equation}
\phi _{\pm }\left( r,\theta \right) \rightarrow \phi _{\mp }\left( r,\pi
-\theta \right) ,  \label{mirror}
\end{equation}%
gives rise to an SV which is a mirror image of (\ref{ansatz}), with the
winding numbers $\left( m_{+},m_{-}\right) =$ $\left( 0,1\right) $ replaced
by $\left( m_{+},m_{-}\right) =\left( -1,0\right) $:%
\begin{equation}
\phi _{+}\left( x,y,t\right) =-e^{-i\mu t-i\theta }r\,f_{2}(r^{2}),~\phi
_{-}=e^{-i\mu t}f_{1}(r^{2}).  \tag{9.13}
\end{equation}%
The coexistence of the two species of mutually-symmetric vortices, defined
by Eqs. (\ref{ansatz}) and (\ref{mirror}), implies the degeneracy of the GS,
which is possible in nonlinear quantum systems, while being prohibited by
general principles of linear quantum mechanics \cite{Landau}.

\begin{figure}[b]
\begin{center}
\includegraphics[width=0.98\textwidth]{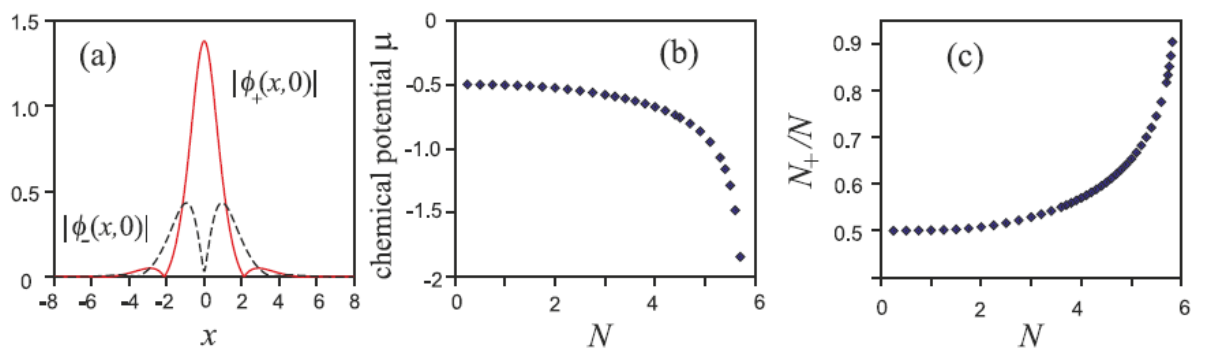}
\end{center}
\caption{(a) Profiles of components $\left\vert \protect\phi %
_{+}(x,0)\right\vert $ and $\left\vert \protect\phi _{-}(x,0)\right\vert $
(as marked near the lines) of a stable SV soliton, shown by their
cross-sections. The soliton is produced as a numerical solution of Eqs. (%
\protect\ref{SOCphi-}) and (\protect\ref{SOCphi+}) with $\protect\lambda %
_{D}=\Omega =\protect\gamma =0$. The total norm (\protect\ref{Nsoc}) is $N=5$%
. (b) Chemical potential $\protect\mu $ of the SV solitons vs. norm $N$ for
the same parameters. (c) The share of the fundamental component of the SV
soliton in its total norm, $N_{+}/N$, as a function of $N$. The figure is
borrowed from Ref. \protect\cite{Ben Li}.}
\label{fig2}
\end{figure}

The localized modes of the SV type exist at values of the chemical potential
$\mu <\mu _{\mathrm{edge}}\equiv -\lambda ^{2}/2$, which is determined by
the edge of the semi-infinite bandgap in the case of $\lambda _{D}=\Omega =0$%
, see Eq. (\ref{semi-inf}). They were produced as stationary numerical
solutions of Eqs. (\ref{SOCphi-}) and (\ref{SOCphi+}), by means of
imaginary-time simulations, and also by means of VA, giving results which
are very close to the numerically found ones \cite{Ben Li}. A typical
example of a stable SV soliton is displayed in Fig. \ref{fig2}(a).

The full family of the SV solitons is represented by the respective $\mu (N)$
dependence in Fig. \ref{fig2}(b), which satisfies the VK criterion (\ref%
{dN/dmu}). Systematic simulations of the perturbed evolution demonstrate
that the entire family is \emph{completely stable} at $\gamma <1$ in Eqs. (%
\ref{SOCphi-}) and (\ref{SOCphi+}) (the case of $\gamma >1$ is considered
below). Further, Fig. \ref{fig2}(c) demonstrates that the fundamental
(zero-vorticity) component is always a dominant one as concerns the
distribution of the total norm between the two components. In the limit of $%
\mu \rightarrow -\infty $, norm $N_{-}$ vanishes, i.e., the vortex component
becomes empty, while the fundamental one degenerates into the
single-component TS with norm (\ref{Pcr}). In fact, the full stability of
the SV family is provided, as mentioned above, by the fact that at all
finite values $\mu <0$ the total norm is smaller than the critical one (\ref%
{Pcr}), hence the solitons are secured against the onset of the critical
collapse \cite{Ben Li}. It is also relevant to stress that, as Fig. \ref%
{fig2}(b) demonstrates, the family of SVs extends up to $N\rightarrow 0$, so
that there is no minimum (threshold) norm necessary for their existence.

\subsubsection{MM solitons}

\begin{figure}[b]
\begin{center}
\includegraphics[width=0.98\textwidth]{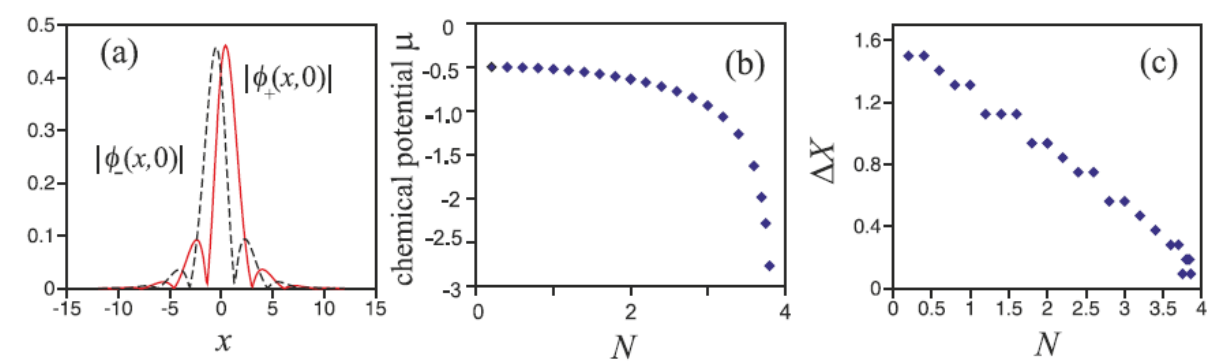}
\end{center}
\caption{(a) Profiles of components $\left\vert \protect\phi %
_{+}(x,0)\right\vert $ and $\left\vert \protect\phi _{-}(x,0)\right\vert $
(as marked near the lines) of a stable MM soliton, shown by means of their
cross-sections, for parameters $\protect\lambda _{D}=\Omega =0$, $\protect%
\gamma =2$ in Eqs. (\protect\ref{SOCphi-}) and (\protect\ref{SOCphi+}). The
total norm of the soliton is $N=3$. (b) The chemical potential $\protect\mu $
of the stable MM vs. its norm $N$ for the same parameters. (c) Separation $%
\Delta \,X$ between peak positions of $|\protect\phi _{+}(x,0)|$ and $|%
\protect\phi _{-}(x,0)|$ vs. $N$, for the same parameters. The figure is
borrowed from Ref. \protect\cite{Ben Li}. \ \ \ \ \ \ \ }
\label{fig3}
\end{figure}

Another type of 2D localized states can be produced by the imaginary-time
simulations of Eqs. (\ref{SOCphi-}) and (\ref{SOCphi+}) with the following
input, which may also serve as the variational \textit{ansatz}:
\begin{equation}
\left( \phi _{+}^{0}\right) _{\mathrm{MM}}=B_{1}e^{-\beta
_{1}r^{2}}-B_{2}r\,e^{-i\theta -\beta _{2}r^{2}},\quad \left( \phi
_{-}^{0}\right) _{\mathrm{MM}}=B_{1}\,e^{-\beta
_{1}r^{2}}+B_{2}r\,e^{i\theta -\beta _{2}r^{2}}.  \label{MM ansatz}
\end{equation}%
Unlike ansatz (\ref{ansatz}) for the SV soliton, the one (\ref{MM ansatz})
is not compatible with the underlying equations, but it is sufficient to
generate a family of numerical solutions for mixed-mode (MM) solitons. They
are built, roughly speaking, as superpositions of SV-like states and their
mirror image given by Eq. (\ref{mirror}) (the MM state is converted into
itself by this transformation). The name of MM stems from the fact that
input (\ref{MM ansatz}) mixes zero-vorticity and vortical terms in each
component. Typical cross-section profiles of the numerically found MM's
components are presented in Fig. \ref{fig3}(a), a characteristic feature
being separation between maxima of the two components (the dependence of the
separation on the total norm is plotted in Fig. \ref{fig3}(c)).

The MM soliton family, similar to its SV counterpart, is characterized in
Fig. \ref{fig3}(b) by the respective dependence $\mu (N)$, cf. Fig. \ref%
{fig2}(b). It also satisfies the VK criterion (\ref{dN/dmu}), and does not
require any minimum value of the norm for the existence of the MM solitons.
In the limit of $\mu \rightarrow \infty $, the MM degenerates into a
symmetric state with identical TSs in the two components (i.e., the vortex
terms vanish in this limit). The corresponding limit value of the norm is
determined by the TS value (\ref{Pcr}), rescaled due the presence of the
cross-attraction: $N_{\mathrm{MM}}\left( \mu \rightarrow -\infty \right)
=2N_{\mathrm{cr}}/(1+\gamma )$.

\begin{figure}[b]
\begin{center}
\includegraphics[width=0.58\textwidth]{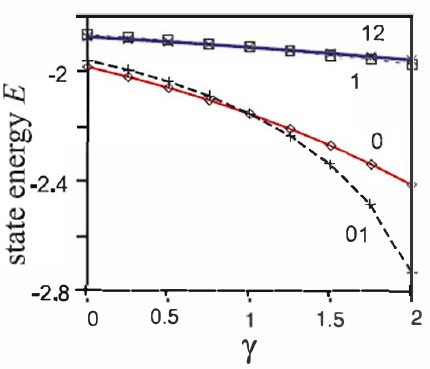}
\end{center}
\caption{Curves labeled by $0$ and $01$ represent, severally, energy (%
\protect\ref{EEEEE}) of the SV and MM solitons vs. the cross-attraction
coefficient $\protect\gamma $, produced by the stationary solution of Eqs. (%
\protect\ref{SOCphi-}) and (\protect\ref{SOCphi+}) with $\protect\lambda %
_{D}=\Omega =0$, at a fixed value of the total norm, $N=3.7$. Additional
(higher) curves represent the energy for excited states (not discussed\
here, as they are completely unstable), see further details in Ref. \protect
\cite{Ben Li}.}
\label{fig4}
\end{figure}

The MMs are unstable if the cross-nonlinear coefficient $\gamma $ in Eqs. (%
\ref{SOCphi-}) and (\ref{SOCphi+}) takes values $\gamma <1$ (recall that the
VK criterion is necessary but not sufficient for the stability). On the
other hand, in the case of $\gamma >1$ (i.e., if the cross-attraction is
stronger than the self-attraction) the SV solitons are unstable and lose the
role of the GS, while the MMs become stable, becoming the system's GS. These
findings may be naturally explained by calculation of energy (\ref{EEEEE})
for the SV and MM solitons with equal values of the total norm. A typical
result is presented in Fig. \ref{fig4}. It is seen that the SV and MM
realize the GS (the energy minimum) at $\gamma <1$ and $\gamma >1$,
respectively. The switch of the GS at $\gamma =1$ from SV to MM is not
surprising, as it corresponds to the special case of the \textit{Manakov's
nonlinearity} \cite{Manakov}, with equal strengths of the inter- and
intra-component attraction, that may readily lead to various degeneracies.

Further numerical analysis demonstrates that unstable SV modes (at $\gamma >1
$) develop spontaneous motion with a nearly constant velocity, keeping their
initial shape intact. On the other hand, the instability of the MM solitons
at $\gamma <1$ leads to their spontaneous rearrangement into stable SV
states \cite{Ben Li}.

\subsubsection{Mobility of the SV and MM solitons}

The mobility of the SV and MM solitons is a nontrivial issue because the SOC
terms in Eqs. (\ref{SOCphi-}) and (\ref{SOCphi+}) break the Galilean
invariance of the system, i.e., solutions for solitons moving steadily at
velocity $\mathbf{v}=(v_{x},v_{y})$ \emph{cannot} be produced by the formal
Galilean boost,%
\begin{equation}
\phi _{\pm }\left( \mathbf{r},t\right) \equiv \tilde{\phi}_{\pm }(\mathbf{r-v%
}t,t)\exp \left( i\mathbf{v}\cdot \mathbf{r}-\frac{i}{2}v^{2}t\right) .
\label{boost}
\end{equation}%
To address the mobility problem, Eqs. (\ref{SOCphi-}) and (\ref{SOCphi+})
(with $\lambda _{D}=\Omega =0$) can be rewritten, for the wave function $%
\phi _{\pm }=\phi _{\pm }(\mathbf{r-v}t,t)$ in the moving reference frame,
with $\mathbf{r}$ replaced by $\mathbf{r-v}t$:
\begin{equation}
i\frac{\partial \phi _{\pm }}{\partial t}-i\left( \mathbf{v}\cdot \nabla
\right) \phi _{\pm }=-\frac{1}{2}\nabla ^{2}\phi _{\pm }-(|\phi _{\pm
}|^{2}+\gamma |\phi _{\mp }|^{2})\phi _{\pm }\pm \lambda \widehat{D}^{[\mp
]}\phi _{\mp },  \label{moving+}
\end{equation}%
In particular, in the case of $v_{x}=0$ the formal Galilean boost (\ref%
{boost}) casts Eq. (\ref{moving+}) in the form that includes terms
representing the linear \textit{Rabi mixing} of the two components, with an
effective coefficient $\lambda v_{y}$:
\begin{equation}
i\frac{\partial \tilde{\phi}_{\pm }}{\partial t}=-\frac{1}{2}\nabla _{\pm
}^{2}\tilde{\phi}_{\pm }-(|\tilde{\phi}_{\pm }|^{2}+\gamma |\tilde{\phi}%
_{\mp }|^{2})\tilde{\phi}_{\pm }\pm \lambda \widehat{D}_{\mp }^{[\mp ]}%
\tilde{\phi}_{\mp }+\lambda v_{y}\tilde{\phi}_{\mp }~.  \label{Rabi}
\end{equation}%
A straightforward impact of the Rabi terms in Eq. (\ref{Rabi}) is a shift
of the edge of the semi-infinite bandgap from the above-mentioned value $\mu
_{\mathrm{edge}}=-\lambda ^{2}/2$ to $\mu _{\mathrm{edge}}=-\left( \lambda
^{2}/2+\left\vert \lambda v_{y}\right\vert \right) $.

The imaginary-time evolution method applied to Eq. (\ref{Rabi}) produces
steady-state solutions in the form of moving MM solitons in a finite
interval of velocities. In particular, for $\gamma =2$, when the GS with $%
\mathbf{v}=0$ is represented by the MM (as said above), the existence
interval for the moving MMs with fixed norm $N=3.1$ and $v_{x}=0$ is $%
\left\vert v_{y}\right\vert <\left( v_{y}\right) _{\max }^{(\mathrm{MM}%
)}\approx 1.8$. The MM's amplitude monotonously decreases with the increase
of $v_{y}$ and vanishes at $v_{y}=\left( v_{y}\right) _{\max }^{(\mathrm{MM}%
)}$ \cite{Ben Li}.

The availability of the stably moving MMs with velocities $\pm v_{y}$
suggests a possibility to simulate collisions between them. The result in
fusion of the solitons into a single mode of the MM type, which is
spontaneously drifting along the $x$ direction \cite{Ben Li}.

At $\gamma <1$, when, as said above, the SV is the GS for $\mathbf{v}=0$,
the numerical solution of Eqs. (\ref{Rabi}) for the moving solitons tends to
converge not to an SV, but rather to an MM state, which turns out to be
stable. Thus, moving SVs are fragile objects, while the MMs are, on the
contrary, robust in the state of motion. This difference is explained by the
fact that the Rabi-mixing terms in Eq. (\ref{Rabi}) do not allow the two
components to maintain different winding numbers, $0$ and $1$, which are the
hallmark of the SV states \cite{Ben Li}.

\subsection{3D metastable solitons in the SOC system}

As said above, 2D solitons in the SOC system can be stabilized because the
SOC terms make it possible to create the solitons with the norm falling
below the threshold (critical) value $N_{\mathrm{cr}}$ necessary for the
onset of the critical collapse. The difficulty in the use of the SOC for the
stabilization of 3D solitons is that, on the contrary to the 2D situation,
the supercritical collapse in 3D has zero threshold \cite{Berge',Fibich}.
Nevertheless, it was demonstrated by means of the VA and numerical methods
that the SOC terms can support \textit{metastable} 3D solitons in the
self-attractive binary condensate, even though such a system has no GS (in
other words, the energy is unbounded from below) \cite{Han Pu 3D}. The
metastability implies stability against small perturbations, while suddenly
applied strong compression may provoke the onset of the supercritical
collapse.

An appropriate 3D GP equation for the pseudo-spinor wave function $\Psi
=\left( \psi _{+},\psi _{-}\right) $ includes the SOC\ of the \textit{Weyl
type} \cite{Weyl} with real coefficient $\lambda $ \cite{Han Pu 3D}:
\begin{equation}
\left[ i\frac{\partial }{\partial t}+\frac{1}{2}\nabla ^{2}+i\lambda \nabla
\cdot \mathbf{\sigma }+\left(
\begin{array}{cc}
|\psi _{+}|^{2}+\eta |\psi _{-}|^{2} & 0 \\
0 & |\psi _{-}|^{2}+\eta |\psi _{+}|^{2}%
\end{array}%
\right) \right] \Psi =0,  \label{3D SOC}
\end{equation}%
where $\mathbf{\sigma =}\left( \sigma _{x},\sigma _{y},\sigma _{z}\right) $
is the vector of the Pauli matrices, and cross-attraction coefficient $\eta $
plays the same role as $\gamma $ in Eqs. (\ref{SOCphi-}) and (\ref{SOCphi+}%
). Stationary soliton states of the SV and MM types were predicted by the
application of VA to Eq. (\ref{3D SOC}), and then produced in a numerical
form. The solutions are characterized by their total norm, defined as the 3D
version of Eq. (\ref{Nsoc}).

\begin{figure}[tbh]
\includegraphics[width=1\columnwidth]{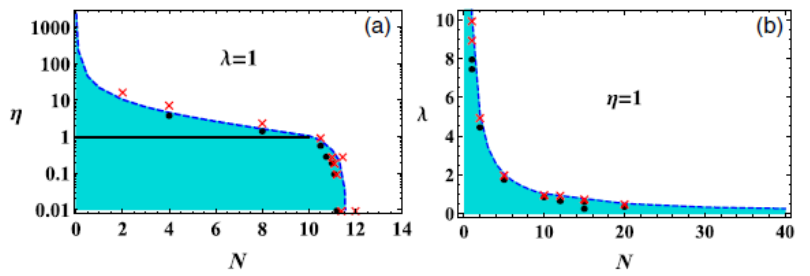}
\caption{3D metastable solitons, produced by Eq. (\protect\ref{3D SOC}), are
predicted by the VA in blue shaded regions of the respective parameter
planes. In (a), these are SVs at $\protect\eta <1$, and MMs at $\protect\eta %
>1$, with the boundary between them marked by the black solid line. In (b),
the entire existence area is filled by the solitons of both types, as the
SVs and MMs have equal energies at $\protect\eta =1$. The predictions are
confirmed by numerical simulations, as indicated by red crosses and black
circles, which indicate, respectively, the absence and presence of stable
soliton solutions for respective sets of parameters. The figure is borrowed
from Ref. \protect\cite{Han Pu 3D}.}
\label{fig5}
\end{figure}

Results are summarized in Fig.~\ref{fig5}, in which stable 3D solitons are
predicted by the VA to exist in the shaded areas. In particular, an
important conclusion is that, for fixed $\lambda $ and $\eta $ in Eq. (\ref%
{3D SOC}), the 3D solitons exist in a finite interval of the norm, $0\leq
N\leq N_{\max }\left( \lambda ,\eta \right) $. As well as in the 2D SOC
system, states with the lowest energy are predicted to be SV and MM, in the
cases of $\eta <1$ and $\eta >1$, respectively.

The VA\ prediction for the existence of the metastable 3D solitons was
verified by simulations of Eq. (\ref{3D SOC}). To this end, stationary
states were generated by imaginary-time simulations. Then, their stability
was tested by means of real-time simulations with small random perturbations
added to the input. The results are shown by red crosses and black circles
in Fig. \ref{fig4}. It is seen that the numerically identified stability
boundary is in good agreement with the VA prediction. Typical examples of
the 3D metastable solitons of the SV and MM types are displayed in Fig. \ref%
{fig6}.

The mobility of the 3D solitons was addressed too, with a conclusion that,
similar to the 2D case, steady motion is possible up to a certain maximum
velocity, at which the amplitude of the soliton vanishes
\begin{figure}[tbh]
\includegraphics[width=0.75\columnwidth]{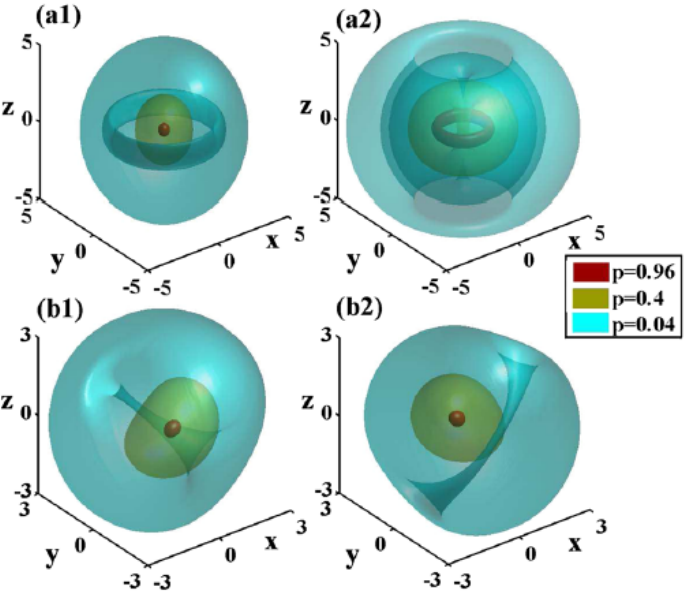}
\caption{Density profiles of metastable 3D solitons with norm $N=8,$
produced by the numerical solution of Eq. (\protect\ref{3D SOC}). (a) An SV
soliton for $\protect\eta =0.3$, whose zero-vorticity and vortical
components, $|\protect\psi _{+}|$ and $|\protect\psi _{-}|$, are plotted in
(a1) and (a2), respectively. (b) An MM soliton for $\protect\eta =1.5$, with
(b1), (b2) displaying $|\protect\psi _{+}|$ and $|\protect\psi _{-}|$,
respectively. In each subplot, different colors represent constant-magnitude
surfaces, $|\protect\psi _{\pm }|=\left( 0.96,0.4,0.04\right) \times |%
\protect\psi _{\pm }|_{\mathrm{max}}$. The figure is borrowed from Ref.
\protect\cite{Han Pu 3D}.}
\label{fig6}
\end{figure}

\section{Emulation of the spin-orbit coupling (SOC) in optical systems}

The similarity between GP equations for matter waves and NLS equations in
optics suggests that many phenomena from the realm of BEC\ may be emulated
in optics, including SOC \cite{Bliokh SOC in opt review,Shenhe}. In
particular, it is possible to elaborate a counterpart for the matter-waves
SOC in 2D systems in terms of the light propagation in dual-core planar
optical waveguides (\textit{couplers}), with amplitudes of the
electromagnetic waves in the two cores emulating two components of the BEC\
pseudo-spinor wave function. Assuming that each core carries the intrinsic
cubic self-focusing, it is possible to design optical setups which maintain
\emph{stable} 2D optical solitons in the spatiotemporal domain \cite{SOC
emulation Barcelona,PT-Sakaguchi}, in spite of the occurrence of the
critical collapse in the same systems.

A crucially important feature of the SOC scheme in the binary BEC, which
makes it possible to achieve the stabilization of the solitons, is that the
linear mixing between the two components of the wave function in Eqs. (\ref%
{SOCphi-}) and (\ref{SOCphi+}) is mediated by the terms with the first-order
spatial derivatives. An optical system, which demonstrates the same feature,
was elaborated in Ref. \cite{SOC emulation Barcelona} in terms of the
spatiotemporal propagation of light in a planar dual-core coupler. As
mentioned above, amplitudes of the electromagnetic waves in the parallel
cores, $q_{1}$ and $q_{2}$, emulate the two components of the pseudo-spinor
wave function in the SOC BEC. In this case, the counterparts of the first
spatial derivatives in Eqs. (\ref{SOCphi-}) and (\ref{SOCphi+}) are time
derivatives, which represent the temporal dispersion of the inter-core
coupling. The respective system of NLS equation, written in the scaled form,
is

\begin{equation}
i\frac{\partial q_{1}}{\partial \xi }+\frac{1}{2}\left( \frac{\partial
^{2}q_{1}}{\partial \eta ^{2}}+\frac{\partial ^{2}q_{1}}{\partial \tau ^{2}}%
\right) -\left\vert q_{1}\right\vert ^{2}q_{1}-\left( c+i\delta \frac{%
\partial }{\partial \tau }\right) q_{2}-\beta q_{1},  \label{q1}
\end{equation}%
\begin{equation}
i\frac{\partial q_{2}}{\partial \xi }+\frac{1}{2}\left( \frac{\partial
^{2}q_{2}}{\partial \eta ^{2}}+\frac{\partial ^{2}q_{2}}{\partial \tau ^{2}}%
\right) -\left\vert q_{2}\right\vert ^{2}q_{2}-\left( c+i\delta \frac{%
\partial }{\partial \tau }\right) q_{1}+\beta q_{2},  \label{q2}
\end{equation}%
where $\xi $ and $\eta $ are the propagation distance and transverse
coordinate in the planar waveguide, $\tau $ is the temporal variable, while
terms $\sim \partial ^{2}/\partial \eta ^{2}$ and $\partial ^{2}/\partial
\tau ^{2}$ represent, respectively,\ the paraxial diffraction and anomalous
GVD. Further, real $c$ is the inter-core coupling, real $\delta $ accounts
for the temporal dispersion of the coupling \cite{Chiang}, and $\beta $,
which is the counterpart of the ZS coefficient $\Omega $ in Eqs. (\ref%
{SOCphi-}), represents possible asymmetry (refractive-index mismatch)
between the optical cores. The model includes the usual intra-core Kerr
nonlinearity, whose coefficient scaled to be $1$. The SOC is emulated by the
combination of the terms $\sim \delta $ and $\beta $.

Plane-wave solutions of the linearized version of Eqs. (\ref{q1}) and (\ref%
{q2}) are looked for as $q_{1,2}\sim \exp \left( ik\xi -i\omega \tau
+bz\right) $, where $b$ is the propagation constant, $k$ the transverse
wavenumber, and $\omega $ the excitation frequency. The resulting dispersion
relation is
\begin{equation}
b_{\pm }=-\frac{1}{2}\left( k^{2}+\omega ^{2}\right) \pm \sqrt{\beta
^{2}+(c+\delta \cdot \omega )^{2}},  \label{b}
\end{equation}%
cf. Eq. (\ref{SOC spectrum}). Stationary soliton solutions of Eqs. (\ref{q1}%
) and (\ref{q2}) are looked for as%
\begin{equation}
q_{1,2}=w_{1,2}(\tau ,\eta )\exp \left( ibz\right) ,  \label{q-soliton}
\end{equation}%
where $w_{1,2}$ are complex functions whose real parts are even with respect
to $\eta $ and $\tau $, while the imaginary ones are even in $\eta $ but odd
in $\tau $. A typical example of a stable soliton is displayed in Fig. \ref%
{fig7}, with a bell-shaped $\left\vert w_{1}\left( \tau ,\eta \right)
\right\vert $ and double-peaked $\left\vert w_{2}\left( \tau ,\eta \right)
\right\vert $. The difference between the shapes of the components is
similar to that between the zero-vorticity and vortex components of the SV
soliton (\ref{ansatz}).
\begin{figure}[tbh]
\includegraphics[width=0.75\columnwidth]{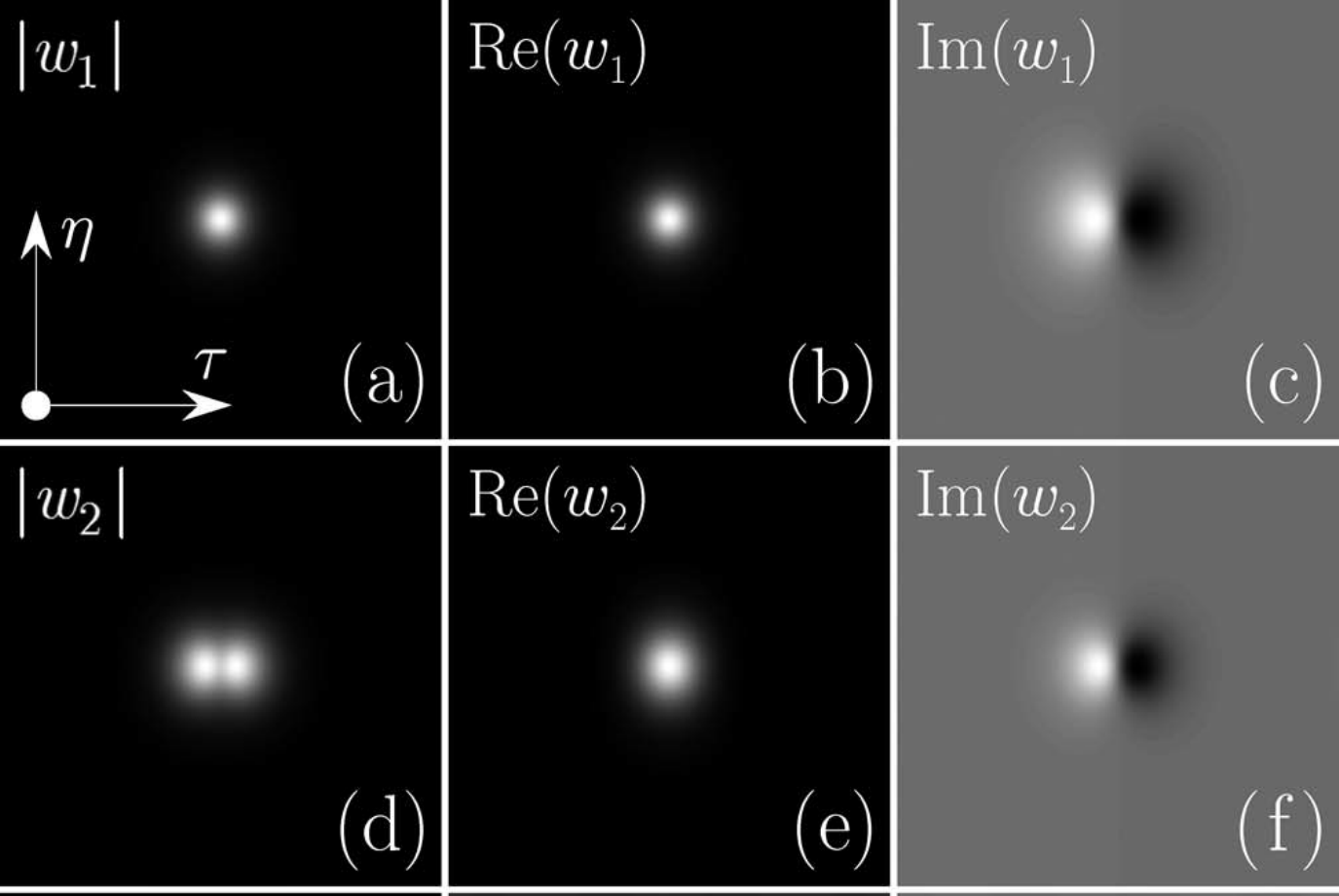}
\caption{An example of a stable spatiotemporal soliton produced by Eqs. (%
\protect\ref{q1}), (\protect\ref{q2}), and (\protect\ref{q-soliton}). The
panels display absolute values of stationary fields $w_{1.2}$ and their real
and imaginary parts. Parameters are $c=\protect\delta =1$, $\protect\beta =3$%
, and $b=1$. The figure is borrowed from Ref. \protect\cite{SOC emulation
Barcelona}.}
\label{fig7}
\end{figure}

\begin{figure}[tbh]
\centering\includegraphics[width=0.75\columnwidth]{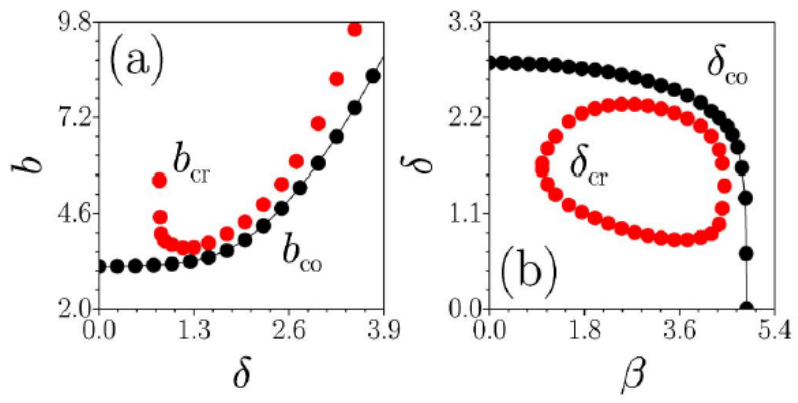}
\caption{Existence and stability domains for the solitons generated by Eqs. (%
\protect\ref{q1}) and (\protect\ref{q2}) in the plane of $\left( \protect%
\delta ,b\right) $ (a), and $\left( \protect\beta ,\protect\delta \right) $
(b). In (a), the solitons exist at $b>b_{\mathrm{co}}$, and are stable at $%
b>b_{\mathrm{cr}}$. In (b), they exist at $\protect\delta >\protect\delta _{%
\mathrm{co}}$, being stable in the domain bounded by the red dots, labeled
as $\protect\delta _{\mathrm{cr}}$. In both plots, $c=1$. The figure is
borrowed from Ref. \protect\cite{SOC emulation Barcelona}.}
\label{fig8}
\end{figure}

Boundaries of the existence and stability domains for the solitons in
parameter planes $\left( \delta ,b\right) $ and $\left( \beta ,\delta
\right) $ are displayed in Fig. \ref{fig8}. The existence (cutoff)
boundaries, $b=b_{\mathrm{co}}$ and $\delta =\delta _{\mathrm{co}}$, which
are marked in the plots, correspond to the edge of the linear spectrum (\ref%
{b}), beyond which the solitons cannot exist. It is seen that the inter-core
coupling dispersion, $\delta $, and inter-core mismatch, $\beta $, are
necessary for the stability of the 2D solitons. This conclusion is natural,
because, as mentioned above, SOC in the dual-core optical system is emulated
by the interplay of these factors, and 2D solitons cannot be stable in the
absence of the SOC.

\section{Quantum droplets}

Among settings which have recently been elaborated for the creation of
multidimensional soliton-like modes, most advanced, as concerns the theory
and experiment alike, is the work with QDs in atomic BEC. Theoretical
predictions for fundamental and topologically structured stable QDs have
been reported in many papers, starting from the pioneering work of Petrov
\cite{Petrov}, who considered a binary condensate with the inter-component
attraction (imposed by means of the Feshbach resonance) slightly exceeding
the intra-component self-repulsion. In the symmetric setting, with equal
wave functions of the two components, the resulting 3D GP equation
demonstrates weak cubic self-attraction. The stabilization against the
supercritical collapse is provided by the correction to the MF approximation
produced by quantum fluctuations around MF states. The correction was
derived in 1957 by Lee, Huang, and Yang (LHY) \cite{LHY}. As demonstrated by
Petrov \cite{Petrov}, in terms of the effective GP equation the LHY
correction is represented by the quartic self-repulsive term, which arrests
the development of the collapse. The balance between the inter-species MF
cubic attraction and LHY-induced quartic repulsion gives rise to stable 3D
and quasi-2D (\textquotedblleft pancake-shaped") localized bound states in
the form of the QDs (see review \cite{Sandy review}). The prediction has
been quickly confirmed by experimental observations of stable quasi-2D \cite%
{Leticia1} and 3D \cite{Inguscio} QDs in condensates of $^{39}$K atoms, with
the binary structure represented by a mixture of two different atomic
states. QDs have also been created in single-component condensates of
magnetic atoms, with long-range attraction provided by interactions between
atomic magnetic moments in ultracold gases of erbium \cite{Santos PRX} and
dysprosium \cite{Pfau-Nature}.

The competition between the MF attraction and LHY repulsion admits atomic
densities which cannot exceed a certain maximum value, thus making the
condensate effectively incompressible. This is a reason why this quantum
state of matter is identified as a superfluid, and localized states filled
by it are called \textquotedblleft droplets".

\subsection{The theoretical models for QDs in three and two dimensions}

The energy density of a binary BEC in the MF\ approximation is
\begin{equation}
\mathcal{E}_{\mathrm{MF}}=\frac{1}{2}g_{11}n_{1}^{2}+\frac{1}{2}%
g_{22}n_{2}^{2}+g_{12}n_{1}n_{2},  \label{energy density}
\end{equation}%
where $g_{11,22}>0$ and $g_{12}<0$ are, respectively, strengths of the
intra-component repulsion and inter-component attraction, and $n_{1,2}\equiv
\left\vert \phi _{1,2}\right\vert ^{2}$ are densities of the two components,
expressed in terms of the respective wave functions $\phi _{1,2}$.

The LHY correction to the MF energy density (\ref{energy density}),
originating from the zero-point energy of Bogoliubov excitations around the
MF state, takes the following form \cite{Petrov}:
\begin{equation}
\mathcal{E}_{\mathrm{LHY}}=\frac{64}{15\sqrt{\pi }}gn^{2}\sqrt{na_{s}^{3}},
\label{LHY corr}
\end{equation}%
where $a_{s}>0$ is the \textit{s}-wave scattering length, $n$ is the density
of both components, assuming that they are equal, $g\equiv 4\pi \hbar
^{2}m/a_{s}$, and $m$ is the atomic mass.

In a dilute condensate, the LHY term (\ref{LHY corr}) with scaling $\sim
n^{5/2}$ is, generally, much smaller than the MF ones $\sim n^{2}$ in Eq. (%
\ref{energy density}). However, when the binary condensate is close to the
equilibrium point, at which the MF self-repulsion is nearly balanced by the
inter-component attraction, i.e., $0<-\delta g\equiv -\left( g_{12}+\sqrt{%
g_{11}g_{22}}\right) \ll g_{11,22}$, the LHY correction is essential. The
resulting system of LHY-amended GP equations for the wave functions of the
3D system can be written in a scaled form as
\begin{equation}
i\frac{\partial \psi _{1}}{\partial t}=-\frac{1}{2}\nabla ^{2}\psi
_{1}+(\left\vert \psi _{1}\right\vert ^{2}+g_{\mathrm{LHY}}\left\vert \psi
_{1}\right\vert ^{3})\psi _{1}-g\left\vert \psi _{2}\right\vert ^{2}\psi
_{1},  \label{Petrov1}
\end{equation}%
\begin{equation}
i\frac{\partial \psi _{2}}{\partial t}=-\frac{1}{2}\nabla ^{2}\psi
_{2}+(\left\vert \psi _{2}\right\vert ^{2}+g_{\mathrm{LHY}}\left\vert \psi
_{2}\right\vert ^{3})\psi _{2}-g\left\vert \psi _{1}\right\vert ^{2}\psi
_{2},  \label{Petrov2}
\end{equation}%
where the quartic self-repulsion terms with coefficient $g_{\mathrm{LHY}}\ $%
represent to the LHY correction to the MF, and $g>0$ is the relative
strength of the MF inter-component attraction.

The application of tight confinement to the condensate in one direction
reduces the dimension from $3$ to $2$. The effective 2D LHY-corrected GP
equation for the symmetric system, with $\phi _{1}=\phi _{2}\equiv \phi $,
was derived in Ref. \cite{Petrov Astrakharchik}. In the scaled form, the 2D
equation is
\begin{equation}
i\frac{\partial \phi }{\partial t}=-\frac{1}{2}\nabla ^{2}\psi +\ln \left(
|\phi |^{2}\right) |\phi |^{2}\phi .  \label{2D}
\end{equation}%
The increase of the local density from $|\phi |^{2}<1$ to $|\phi |^{2}>1$
leads to the change of the sign of the logarithmic factor in Eq. (\ref{2D}%
). As a result, the cubic term is self-attractive at small densities,
initiating the spontaneous formation of QDs, and repulsive at large
densities, arresting the transition to the collapse, thus stabilizing the 2D
QDs.

\subsection{3D vortex QDs: theoretical results}

Solutions of Eqs. (\ref{Petrov1}) and (\ref{Petrov2}) for 3D vortex
solitons, with equal components carrying chemical potential $\mu $ and
integer winding number $m$, are looked for, in cylindrical coordinates $%
\left( \rho ,\theta ,z\right) $, as
\begin{equation}
\psi _{1}=\psi _{2}=u\left( \rho ,z\right) \exp \left( -i\mu t+im\theta
\right) ,  \label{psi1=psi2}
\end{equation}%
where real function $u\left( \rho ,z\right) $ satisfies the equation%
\begin{equation}
\mu u+\frac{1}{2}\left( \frac{\partial ^{2}}{\partial \rho ^{2}}+\frac{1}{%
\rho }\frac{\partial }{\partial \rho }+\frac{\partial ^{2}}{\partial z^{2}}-%
\frac{m^{2}}{\rho ^{2}}\right) u+(g-1)u^{3}-g_{\mathrm{LHY}}u^{4}=0.
\label{u}
\end{equation}

For given $m$, soliton families are characterized, as usual, by dependences
of $\mu $ on the total norm, $N=4\pi \int_{0}^{\infty }\rho d\rho
\int_{-\infty }^{+\infty }dzu^{2}(\rho ,z)$. The families were constructed
by means of a numerical solution \cite{swirling}. The crucially important
problem of their stability was addressed by means of a numerical solution of
the linearized Bogoliubov - de Gennes equations for small perturbations
around the stationary solitons, and the results were verified by means of
direct simulations of Eqs. (\ref{Petrov1}) and (\ref{Petrov2}) for the
perturbed evolution.

\begin{figure}[tbp]
{\includegraphics[scale=0.83]{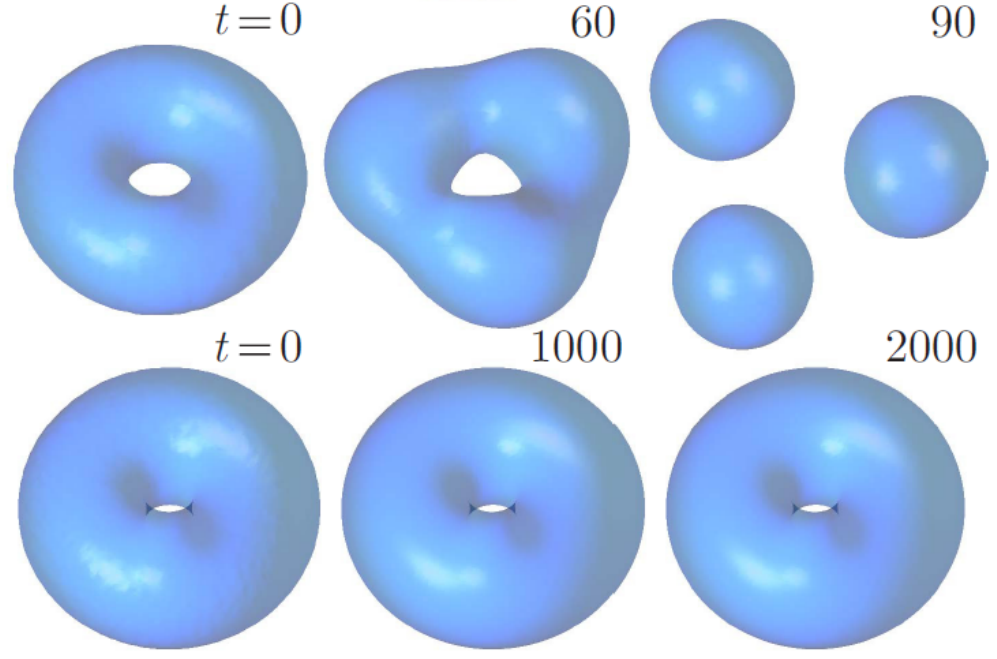}}
\caption{Isosurface density plots, at levels $|\protect\psi _{1,2}|^{2}$ $%
=0.1$ and $|\protect\psi _{1,2}|^{2}$ $=0.5$ in the top and bottom row,
respectively, which demonstrate unstable and stable evolution of 3D
symmetric ($\protect\psi _{1}=\protect\psi _{2}$) vortex QDs with winding
numbers $m_{1,2}=1$, produced by simulations of Eqs. (\protect\ref{Petrov1})
and (\protect\ref{Petrov2}) with $g_{\mathrm{LHY}}=0.50$ and $g=1.75$.
Chemical potentials of the unstable and stable solutions are, respectively, $%
\protect\mu _{1,2}=-0.04$ and $\protect\mu _{1,2}=-0.16$. The figure is
borrowed from Ref. \protect\cite{swirling}.}
\label{fig9}
\end{figure}

The vortex QDs may be stable if their norm is large enough, which is a
generic property of solitons supported by competing nonlinearities \cite%
{Quiroga,3D vort sol CQ,Pego}. In turn, the QDs with a large norm are very
broad, with a \textit{flat-top }(FT)\textit{\ }shape, see examples in Fig. %
\ref{fig9}. In this limit, one may asymptotically neglect the terms $\sim
\rho ^{-1}$ and $\rho ^{-2}$ in Eq. (\ref{u}), as well as $\partial
^{2}u/\partial z^{2}$, reducing Eq. (\ref{u}) to the quasi-1D form,%
\begin{equation}
\mu u+\frac{1}{2}\frac{\partial ^{2}u}{\partial \rho ^{2}}+(g-1)u^{3}-g_{%
\mathrm{LHY}}u^{4}=0.  \label{flat}
\end{equation}%
This equation seems as a formal equation of motion for a particle with
coordinate $u$ in time $\rho $, which conserves the Hamiltonian,%
\begin{equation}
\mathcal{H}=\frac{1}{4}\left( \frac{\partial u}{\partial \rho }\right) ^{2}+%
\frac{\mu }{2}u^{2}+\frac{1}{4}(g-1)u^{4}-\frac{1}{5}g_{\mathrm{LHY}}u^{5}.
\label{H}
\end{equation}%
QD solutions, vanishing at $\rho \rightarrow \infty $, correspond to $%
\mathcal{H}=0$. Finally, the FT solitons asymptotically correspond to
setting $\partial ^{2}u/\partial \rho ^{2}=\partial u/\partial \rho =0$ in
Eqs. (\ref{flat}) and (\ref{H}) (with $\mathcal{H}=0$), which yields
algebraic equations for the chemical potential and density $u^{2}$ of the
asymptotically FT state inside of the broad solitons. The resulting solution
is%
\begin{equation}
\mu _{\mathrm{FT}}=-\frac{25}{216}(g-1)\left( g_{\mathrm{LHY}}\right)
^{-2},~\left( u_{\mathrm{FT}}\right) ^{2}=\frac{25}{36}(g-1)^{2}\left( g_{%
\mathrm{LHY}}\right) ^{-2}~.  \label{FT}
\end{equation}%
The QDs exist in interval $0<-\mu <-\mu _{\mathrm{FT}}$, with $N(\mu =\mu _{%
\mathrm{FT}})=\infty $ (the divergence of the norm at $\mu \rightarrow \mu _{%
\mathrm{FT}}$ is a common feature of NLS equations with competing
nonlinearities \cite{Pushkarovs}). Further, $\left( u_{\mathrm{FT}}\right)
^{2}$ is the largest density which these solutions may attain. As said
above, the latter fact implies that the QDs are filled by the effectively
incompressible quantum fluid. In the experimental realization of QDs, the
density is $\sim 10^{-8}$ g/cm$^{3}$ \cite{Leticia1,Inguscio}, which is five
orders of magnitude lower than the normal air density. In fact, the
superfluid filling QDs is the most rarefied fluid known in nature \cite%
{Petrov}.

The results produced for 3D\ QDs \cite{swirling} are summarized in Fig. \ref%
{fig10}, where panel (a) represents QD families with $m=0,1$ and $2$ by
means of the respective $N(\mu )$ curves, and (b) displays existence and
stability boundaries ($N=N_{\min }$ and $N=N_{\mathrm{st}}$, respectively)
for the family with $m=1$ in the $\left( g,N\right) $ plane. The fact that
the 3D QDs exist above a minimum (threshold) value of $N$ is generic for 3D
models with competing self-attractive and repulsive nonlinearities \cite{3D
vort sol CQ}. In the area $N_{\min }<N<N_{\mathrm{st}}$ in Fig. \ref{fig9}%
(b) the vortical QDs are unstable against spontaneous splitting, as shown in
the top part of Fig. \ref{fig9}. The stability-boundary value $N_{\mathrm{st}%
}$ rapidly increases with the decrease of $g$, and diverges at $g=g_{\min
}\approx 1.3$, all QDs with vorticity $m=1$ being unstable at $g<g_{\min }$.

Although the branch of the QD vortex states with $m=2$ is completely
unstable in the intervals of $N$ and $\mu $ which are shown in Fig. \ref%
{fig10}(a), these solutions become stable at still larger values of $N$ and $%
-\mu $. For instance, the double-vortex QD is stable, with the same values $%
g=0.5$ and $g_{\mathrm{LHY}}=1.75$ as fixed in \ref{fig9}(a), for $\mu
=-0.183$ \cite{swirling}. Plausibly, QDs with $m\geq 3$ and extremely large
norms may be stable too, although they have not been found, as yet. This
fact is explained by scaling
\begin{equation}
N_{\mathrm{st}}\sim m^{6}  \label{m^6}
\end{equation}%
\ for the lowest norm necessary for the stability of the vortex QDs in the
3D setting \cite{2D vortex QD}. Indeed, this steep scaling makes it
difficult to create stable QDs with $m\geq 3$.

It is also possible to consider the QD solutions of Eqs. (\ref{Petrov1}) and
(\ref{Petrov2}) with \textit{hidden vorticity} (HV, i.e., opposite winding
numbers in the two components and zero total angular momentum). They are
defined, according to Ref. \cite{Brtka}, as
\begin{equation}
\psi _{1}=u\left( \rho ,z\right) \exp \left( -i\mu t+im\theta \right) ,~\psi
_{2}=u\left( \rho ,z\right) \exp \left( -i\mu t-im\theta \right) ,
\label{hidden}
\end{equation}%
where real function $u(\rho ,z)$ satisfies the same equation (\ref{u}) as
above. However, unlike the QDs with explicit vorticity, considered above,
the HV ones are completely unstable against spontaneous splitting \cite%
{swirling}.
\begin{figure}[tbp]
\begin{center}
\subfigure[]{\includegraphics[width=0.48\textwidth]{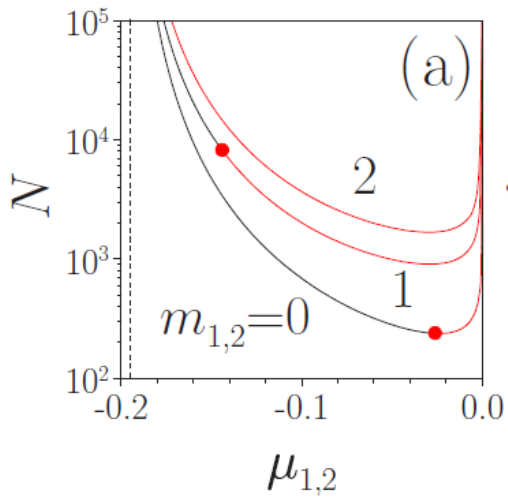}} %
\subfigure[]{\includegraphics[width=0.48\textwidth]{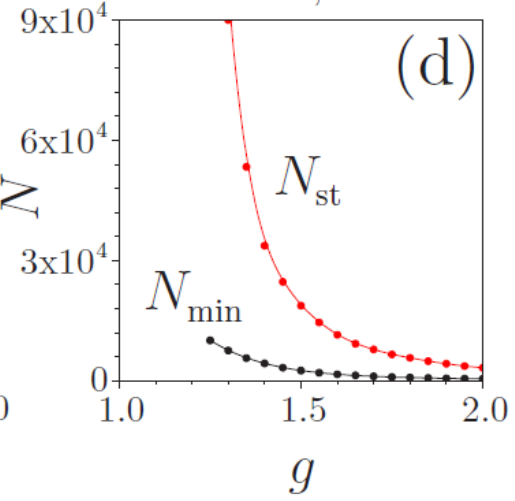}}
\end{center}
\caption{(a) The norm versus the chemical potential for QDs with indicated
values of the vorticity of both components, $m_{1}=m_{2}$, and parameters $%
g_{\mathrm{LHY}}=0.5$ and $g=1.75$ in Eqs. (\protect\ref{Petrov1}) and (%
\protect\ref{Petrov2}). Red and black branches designate unstable and stable
QDs, respectively, with red dots separating stable and unstable segments.
For $m_{1,2}=2$, a stability segment exists too, but it is located at large
values of $N$, outside of the region displayed in this panel. The dashed
vertical line designates the value $\protect\mu _{\mathrm{FT}}$, given by
Eq. (\protect\ref{FT})), at which the QD's norm diverges. (b) The bottom
curve: the minimum norm, $N_{\min }$, above which Eq. (\protect\ref{u})
produces the vortex-QD solutions with winding number $m=1$, as a function of
the attraction strength $g$. The top curve: the critical norm, $N_{\mathrm{st%
}}$, above which the vortex QD is stable, vs. $g$. All the vortex states are
unstable against spontaneous splitting at $g<g_{\min }\approx 1.3$ ($N_{%
\mathrm{st}}$ diverges at $g\rightarrow g_{\min }$). In this panel, $g_{%
\mathrm{LHY}}=0.5$ is fixed. The figure is borrowed from Ref. \cite{swirling}.}
\label{fig10}
\end{figure}

\subsection{2D vortex QDs}

Solutions for 2D QDs, in the form of $\phi =u\left( r\right) \exp \left(
-i\mu t+iS\theta \right) $, and their stability were explored, on the basis
of Eq. (\ref{2D}), in Ref. \cite{2D vortex QD}. Here, $\left( r,\theta
\right) $ are the polar coordinates, the integer vorticity is denoted as $S$%
, and real function $u(r)$ satisfies the equation%
\begin{equation}
\mu u+\frac{1}{2}\left( \frac{d^{2}}{dr^{2}}+\frac{1}{r}\frac{d}{dr}-\frac{%
S^{2}}{r^{2}}\right) u+\ln \left( u^{2}\right) \cdot u^{3}=0,  \label{u2D}
\end{equation}%
cf. Eq. (\ref{u}). The limit (quasi-1D) form of Eq. (\ref{u2D}),
corresponding to dropping terms $\sim r^{-1}$ and $r^{-2}$ (cf. Eq. (\ref%
{flat})), takes the form of a formal equation of motion with Hamiltonian
\begin{equation}
\mathcal{H}^{\mathrm{(2D)}}=\frac{1}{4}\left( \frac{du}{dr}\right) ^{2}+%
\frac{\mu }{2}u^{2}-\frac{1}{4}\ln \left( \frac{u^{2}}{\sqrt{e}}\right) ,
\label{2DH}
\end{equation}%
cf. Eq. (\ref{H}). Similar to Eq. (\ref{FT}) in the 3D case, the quasi-1D
reduction of Eq. (\ref{u2D}) and (\ref{2DH}) predict the chemical potential
and density of the 2D\ FT state (i.e., the largest attainable density for
the 2D superfluid):%
\begin{equation}
\mu _{\mathrm{FT}}^{\mathrm{(2D)}}=-\left( 2\sqrt{e}\right) ^{-1}\approx
-0.3033,~\left( u_{\mathrm{FT}}^{\mathrm{(2D)}}\right) ^{2}=1/\sqrt{e}%
\approx 0.6065.  \label{2DFT}
\end{equation}

\begin{figure}[tbp]
\subfigure[]{\includegraphics[width=0.45\textwidth]{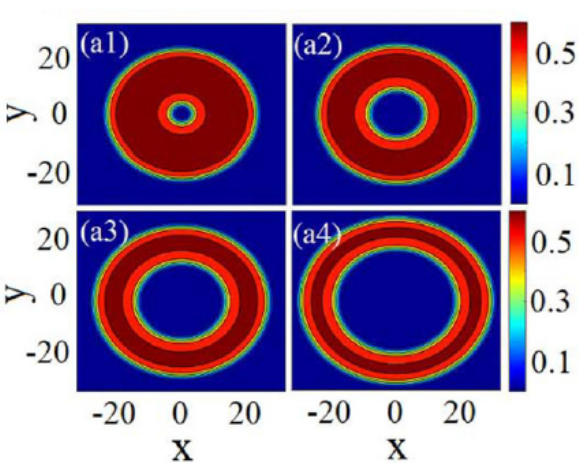}} %
\subfigure[]{\includegraphics[width=0.45\textwidth]{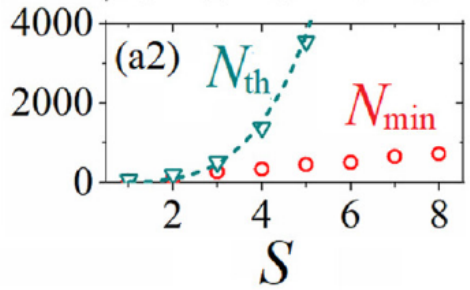}}
\caption{(a) Panels (a1)-(a4) display density patterns of 2D vortex QDs with
$S=1,2,3,4$ and norm $N=1000$, produced by the numerical solution of Eq. (%
\protect\ref{u2D}). The states with $S=1,2,3$ are stable, while the one
with\ $S=4$ is unstable. (b) The minimum norm, $N_{\min }$, necessary for
the existence of the 2D vortex QDs (circles), and the threshold (critical)
value of the norm, $N_{\mathrm{th}}$, above which they are stable
(triangles), vs. vorticity $S$. The results are produced by the numerical
solution of Eq. (\protect\ref{2D}). The dashed curve shows the fit to the
analytically predicted scaling (\protect\ref{S^4}), $N_{\mathrm{th}}=6S^{4}$%
. The figure is borrowed from Ref. \protect\cite{3D vort sol CQ}.}
\label{fig11}
\end{figure}

Numerical results demonstrate that 2D vortex QDs with a sufficiently large
norm, $N=2\pi \int_{0}^{\infty }u^{2}(r)rdr$, may be stable, at least, up to
$S=5$ \cite{2D vortex QD}. Examples are displayed in Fig. \ref{fig11}(a).
The results are summarized in Fig. \ref{fig11}(b) by plotting the lowest
values of the norm which are necessary for the existence and stability of
the 2D vortex QDs, \textit{viz}., $N_{\min }$ and $N_{\mathrm{th}}$,
respectively. An analytical result is that $N_{\mathrm{th}}$ scales as
\begin{equation}
N_{\mathrm{th}}^{\mathrm{(2D)}}\sim S^{4}  \label{S^4}
\end{equation}%
with the increase of $S$ \cite{2D vortex QD}, cf. the steeper scaling in the
3D case, given by Eq. (\ref{m^6}).

QDs of the HV type, corresponding to $m=\pm 1$ in the 2D version of Eq. (\ref%
{hidden}), also have their narrow stability region in the 2D model based on
Eq. (\ref{2D}), with norms $N>N_{\mathrm{th}}^{\mathrm{(HV)}}\simeq 100$
\cite{2D vortex QD}, on the contrary to the full instability of the HV
states in the 3D case, as mentioned above. However, all the 2D HV droplets
with $|m|\geq 2$ are unstable.

A system of 2D GP equations, including both the LHY corrections and SOC
(spin-orbit coupling) between the components, gives rise to vortex QDs of
the MM type, which mix terms with $S=0$ and $+1$ or $-1$ in the two
components (see Eq. (\ref{MM ansatz}) above). These QDs, which feature an
anisotropic shape in 2D, are stable in a certain parameter region \cite{NJP
QD SOC}.

Vortex QDs in a condensate of magnetic atoms, with long-range dipole-dipole
interactions, were theoretically investigated too and found to be \emph{%
completely unstable }(unlike the results presented here) \cite{unstable
vortex}.

\subsection{Experimental observations of QDs}

\begin{figure}[tbp]
{\includegraphics[scale=0.86]{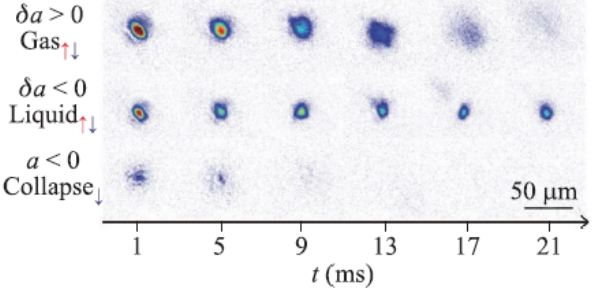}}
\caption{The experimentally observed evolution of the condensates of $^{39}$%
K atoms in the oblate (quasi-2D) setting, displayed by means of density
snapshots taken at different times $t$. Top: Expansion of a the binary
condensate in the absence of the effective MF attraction, with $\protect%
\delta a=1.2a_{0}$ (see Eq. (\protect\ref{delta_a})). Middle: The  formation
of a \emph{stable} oblate QD in the binary condensate with the
inter-component attraction, characterized by $\protect\delta a=-3.2a_{0}<0$.
Bottom: The collapse of a single-component self-attractive condensate, with
scattering length $a=-2.06a_{0}$. The picture is borrowed from Ref.
\protect\cite{Leticia1}. }
\label{fig12}
\end{figure}

The creation of stable QDs with a quasi-2D shape in the mixture of two
different Zeeman states of $^{39}$K atoms (\textquotedblleft up" and
\textquotedblleft down" ones), namely,
\begin{equation}
\left\vert \uparrow \right\rangle \equiv \left\vert F,m_{F}\right\rangle
=\left\vert 1,-1\right\rangle \text{ and }\left\vert \downarrow
\right\rangle =\left\vert 1,0\right\rangle ,  \label{states}
\end{equation}%
where $F$ is the total angular momentum of the atom, and $m_{F}$ is its $z$
component, was first reported in Ref. \cite{Leticia1}. That work used a
setup in which the QDs were self-trapped in the 2D plane, and confined by a
trapping potential in the transverse direction, making the QD shape oblate.
The mixture is characterized by respective scattering length, $a_{\uparrow
\uparrow ,\downarrow \downarrow }>0$ and $a_{\uparrow \downarrow }<0$. The
residual MF interaction was determined by
\begin{equation}
\delta a=a_{\uparrow \downarrow }+\sqrt{a_{\uparrow \uparrow }a_{\downarrow
\downarrow }},  \label{delta_a}
\end{equation}%
adjusted by means of the Feshbach resonance. As shown in Fig. \ref{fig12},
the formation of stable oblate QDs, supported by the balance of MF
attraction and LHY self-repulsion in the binary condensate, was observed in
the case of $\delta a\approx -3.2a_{0}<0$, where $a_{0}\approx 0.0529$ nm is
the Bohr radius. The same figure demonstrates fast decay of the input in the
absence of the attraction ($\delta a>0$), and the collapse of the
single-component self-attractive condensate ($a<0$).

Stable QDs in the 3D isotropic form were created in work \cite{Inguscio},
using the same atomic states (\ref{states}) as in Ref. \cite{Leticia1}, and
adjusting effective interactions by means of the Feshbach resonance. A
holding potential, imposed by three mutually perpendicular red-detuned laser
beams, was employed to prepare the condensate. In the case of $\delta a>0$
in Eq. (\ref{delta_a}) (MF repulsion), the initial state quickly expands
after removing the holding potential, as shown in the top row of Fig. \ref%
{fig13}. On the other hand, the bottom row demonstrates self-trapping of a
\emph{stable} isotropic QD under the action of the MF attraction, with $%
\delta a<0$.

\begin{figure}[tbp]
{\includegraphics[scale=1.08]{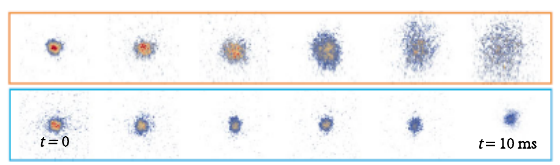}}
\caption{Top: Decay of the initial isotropic input in the absence of the
MF attraction, with $\protect\delta a>0$ in Eq. (\protect\ref{delta_a}).
Bottom: The experimentally demonstrated formation of a \emph{stable}
isotropic QD in the case of the attraction, with $\protect\delta a<0$ in Eq.
(\protect\ref{delta_a})). The horizontal axis represents time varying from $%
t=0$ to $10$ ms. The figure is borrowed from Ref. \protect\cite{Inguscio}.}
\label{fig13}
\end{figure}

Both the oblate and isotropic QDs demonstrate rather short lifetimes,
measured in few tens of milliseconds \cite{Leticia1,Inguscio}. The lifetimes
are limited by three-body collisions, which kick out atoms from the
condensate into a thermal component of the gas.

\begin{figure}[tbp]
\subfigure[]{\includegraphics[width=0.16\textwidth]{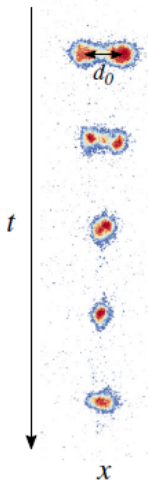}} %
\subfigure[]{\includegraphics[width=0.16\textwidth]{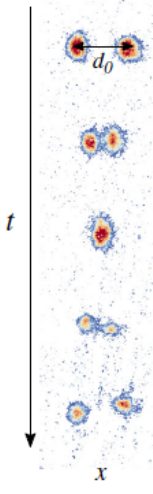}}
\caption{Experimentally observed collisions between moving isotropic QDs:
(a) merger of slowly moving droplets; (b) passage of fast moving ones. The
figure is borrowed from Ref. \protect\cite{Inguscio collision}.}
\label{fig14}
\end{figure}

The stable isotropic QDs can be readily set in motion by kicks, applied to
them by laser pulses. This option suggests a possibility to study collisions
between QDs moving in opposite directions, which was performed
experimentally in Ref. \cite{Inguscio collision}, using the same binary
condensate of $^{39}$K. Note that collisions of droplets of a classical
fluid may lead to their merger into a single one, provided that the surface
tension is sufficiently strong to absorb the kinetic energy of the colliding
pair. Otherwise, the colliding droplets separate in two or more ones \cite%
{classical droplet}. As shown in Fig. \ref{fig14}, similar phenomenology is
exhibited by the colliding QDs: merger of slowly moving droplets, and
quasi-elastic mutual passage of faster ones. Similar results were
demonstrated by simulations of the collisions in the framework of Eqs. (\ref%
{Petrov1}) and (\ref{Petrov2}).

The mechanism stabilizing two-component QDs with MF inter-component
attraction and LHY intra-component self-repulsion applies not only to binary
systems composed of different hyperfine states of the same atom, but also to
heteronuclear mixtures. Experimentally, this possibility was realized in a
mixture of $^{41}$K and $^{87}$Rb atoms \cite{hetero QD}. In particular, stable
QDs were observed with the ratio of atom numbers $N_{\mathrm{K}}/N_{\mathrm{%
Rb}}\approx 0.8$, featuring lifetimes up to $\sim 100$ times longer than in
the monatomic binary condensate.

\section{Conclusion}

This review is focused on selected topics for multidimensional solitons,
which seem most relevant for the ongoing theoretical and experimental work
in this vast area. These are stabilization of 2D and 3D SV and MM solitons
by SOC\ in binary BECs, stabilization of 2D spatiotemporal optical solitons
in dual-core planar waveguides by the photonic counterpart of SOC, and
various theoretical and experimental results for 3D and 2D QDs (soliton-like
states), stabilized by the LHY effect (quantum fluctuations around the MF
states).

There are many other relevant aspects of studies of multidimensional
solitons, which could not be included in this relatively short article. Most
of them are presented in sufficient detail in the recently published book
\cite{book}. In addition to that, there are other significant topics related
to this area, which are presented in other publications. In particular, this
is the propagation and self-trapping of spatiotemporal modes, including ones
carrying the optical angular momentum, in multimode nonlinear optical
fibers, for which the 3D NLS equation is a relevant model \cite{Krupa,OAM in
fibers}. Also related to multidimensional solitons is nonlinear topological
photonics, which addresses the propagation of topologically structured
self-trapped modes in complex media \cite{Bandres,Smirnova}.

The recent progress in the work with multidimensional solitons has been
fast, and there is vast space for further theoretical and experimental
developments in this area.

\section*{Acknowledgments}

I would like to thank my colleagues, with whom I collaborated on topics
addressed in this review: F. Kh. Abdullaev, D. Anderson, G. Astrakharchik,
C. B. de Ara\'{u}jo, B. B. Baizakov, W. B. Cardoso, G. Dong, N. Dror, P. D.
Drummond, A. Gammal, Y. V. Kartashov, V. V. Konotop, H. Leblond, B. Li, M.
Lisak, Z.-H. Luo, Y. Li, D. Mihalache, M. Modugno, W. Pang, M. A. Porras, H.
Pu, J. Qin, A. S. Reyna, H. Sakaguchi, L. Salasnich, M. Salerno, E. Ya.
Sherman, V. Skarka, D. V. Skryabin, L. Tarruell, L. Torner, F. Wise, and L.
G. Wright. I also thank Editors of Advances in Physics X, J. S. Aitchison,
N. Balmforth, and R. Palmer, for their invitation to write this review
article. My work on this topic was partly supported by the Israel Science
Foundation through grant No. 1695/22.

\section*{Disclosure statement}

There is no conflict of interests to declare.

\end{document}